\documentclass[twocolumn]{aastex61}

\shorttitle{Spin-Orbit Misalignments of 3 Planets}
\shortauthors{Johnson et al.}

\begin{document}

\title{Spin-Orbit Misalignments of Three Jovian Planets via Doppler Tomography\footnote{Based in part on observations obtained with the Hobby-Eberly Telescope, which is a joint project of the University of Texas at Austin, the Pennsylvania State University, Stanford University, Ludwig-Maximilians-Universit\"at M\"unchen, and Georg-August-Universit\"at G\"ottingen}} 

\author{Marshall C. Johnson}
\affiliation{Department of Astronomy and McDonald Observatory, University of Texas at Austin, 2515 Speedway, Stop C1400, Austin, TX 78712 USA}
\affiliation{Department of Astronomy, The Ohio State University, 140 West 18$^{\mathrm{th}}$ Ave., Columbus, OH 43210 USA; johnson.7240@osu.edu}

\author{William D. Cochran}
\affiliation{Department of Astronomy and McDonald Observatory, University of Texas at Austin, 2515 Speedway, Stop C1400, Austin, TX 78712 USA}

\author{Brett C. Addison}
\affiliation{Mississippi State University, Department of Physics \& Astronomy, Hilbun Hall, Starkville, MS 39762, USA}

\author{Chris G. Tinney}
\affiliation{Exoplanetary Science Group, School of Physics, University of New South Wales, NSW 2052, Australia}
\affiliation{Australian Centre of Astrobiology, University of New South Wales, NSW 2052, Australia}

\author{Duncan J. Wright}
\affiliation{Exoplanetary Science Group, School of Physics, University of New South Wales, NSW 2052, Australia}
\affiliation{Australian Centre of Astrobiology, University of New South Wales, NSW 2052, Australia}
\affiliation{Australian Astronomical Observatory (AAO), PO Box 915, North Ryde, NSW 1670, Australia}


\begin{abstract}

We present measurements of the spin-orbit misalignments of the hot Jupiters HAT-P-41~b and WASP-79~b, and the aligned warm Jupiter Kepler-448~b. We obtained these measurements with Doppler tomography, where we spectroscopically resolve the line profile perturbation during the transit due to the Rossiter-McLaughlin effect. We analyze time series spectra obtained during portions of five transits of HAT-P-41~b, and find a value of the spin-orbit misalignment of $\lambda = -22.1_{-6.0}^{+0.8 \circ}$. We reanalyze the radial velocity Rossiter-McLaughlin data on WASP-79~b obtained by \cite{Addison13} using Doppler tomographic methodology. We measure $\lambda=-99.1_{-3.9}^{+4.1\circ}$, consistent with but more precise than the value found by \cite{Addison13}. For Kepler-448~b we perform a joint fit to the \emph{Kepler} light curve, Doppler tomographic data, and a radial velocity dataset from \cite{LilloBox15}. We find an approximately aligned orbit ($\lambda=-7.1^{+4.2 \circ}_{-2.8}$), in modest disagreement with the value found by \cite{Bourrier15}. Through analysis of the \emph{Kepler} light curve we measure a stellar rotation period of $P_{\mathrm{rot}}=1.27 \pm 0.11$ days, and use this to argue that the full three-dimensional spin-orbit misalignment is small, $\psi\sim0^{\circ}$.

\end{abstract}


\section{Introduction}

The sky-projected spin-orbit misalignment $\lambda$ is the angle between a planet's orbital angular momentum vector and the rotational angular momentum vector of its host star, as projected onto the plane of the sky. Over the past decade, this angle has been measured for dozens of planets, mostly hot Jupiters \citep[e.g.,][]{Winn05,Triaud10,Addison16}; many of these have highly inclined or even retrograde orbits.

Several patterns have been identified in the distributions of $\lambda$ as a function of various parameters. \cite{Winn10} found that the hot Jupiters with misaligned orbits preferentially occur around stars with $T_{\mathrm{eff}}>6250$ K, and that those around cooler stars are preferentially well-aligned. \cite{Albrecht12} further found that the misaligned planets tend to have longer tidal damping timescales. \cite{Hebrard10} noted that there are no known planets with masses $M_P>3\,M_J$ on retrograde orbits ($90^{\circ}<|\lambda|\leq180^{\circ}$), an effect also attributed to tidal damping of orbital obliquities. \cite{Dawson14} demonstrated that these trends can be explained by the interplay of tides and magnetic braking of rotation for stars on either side of the Kraft break \citep{Kraft67}. The overall scenario is that some fraction of hot Jupiters are initially emplaced on highly misaligned orbits, and the orbital inclinations of planets around slowly-rotating stars are quickly damped out by tides, while those around more rapidly rotating stars (except for the most massive planets) are not. There are, however, still challenges for this scenario, both in terms of theoretical understanding of tides and the extent to which longer-period planets are misaligned \citep{LiWinn16}.

Despite a great deal of theoretical work, the mechanism(s) by which hot Jupiters are emplaced onto misaligned orbits--and even whether an alternate mechanism, rather than tides, might actually be responsible for the trends described above--remains unclear. In general, there are two classes of models for the generation of misalignments--either the misalignments were caused by the same migration processes that brought the hot Jupiters close to their stars, or the planets are misaligned for reasons unrelated to their migration. The first class of models generally consists of various high-eccentricity migration scenarios: planet-planet scattering \citep[e.g.,][]{RasioFord96,Chatterjee08}; the Kozai-Lidov mechanism \citep[e.g.,][]{FabryckyTremaine07,Naoz12}; and chaotic orbital evolution due to planets in mean motion resonances with eccentric, mutually inclined orbits \citep{Barnes15}. Mechanisms that could result in misaligned orbits largely unrelated to planet migration include: migration within a protoplanetary disk misaligned with respect to the star due to accretion of material with time-variable bulk angular momenta \citep[e.g.,][]{Bate10,Fielding15}, gravitational torques from a binary companion or the birth cluster \citep[e.g.,][]{Batygin12,Spalding14}, or magnetic torques from the star \citep[e.g.,][]{Lai11,FoucartLai11}; or, angular momentum transport within hot stars due to internal gravity waves, making the stellar atmosphere not rotate in a manner indicative of the bulk stellar angular momentum \citep{Rogers12,Rogers13}. Alternatives to tides for realignment, both of which would require a misaligned protoplanetary disk, are early ingestion of a hot Jupiter by the star, which could realign less massive stars but not more massive ones \citep{MatsakosKonigl15}, and realignment of the star with the disk due to stellar magnetic torques, which would be more effective for lower-mass stars and their stronger magnetic fields \citep{Spalding14}.

A variety of further observations are needed to distinguish among these models. Since several of the misalignment mechanisms rely upon the presence of additional objects in the system, systematic searches to detect (or set limits upon) additional planetary or stellar companions are necessary and ongoing \citep[e.g.,][]{Knutson14,Ngo16}. Expanding the sample of measured spin-orbit misalignments is also critical. Although many hot Jupiters already have measurements of $\lambda$, more measurements can identify edge cases that help us further probe and constrain the models. An example of this is HATS-14 b, the only known hot Jupiter with a highly inclined orbit around a cool star that neither orbits a young star nor has any additional currently known objects in its system \citep{Zhou15}. Observations of additional classes of planets beyond hot Jupiters are also important, as the different models make different predictions for other populations. Spin-orbit misalignments have been measured for a few small planets \citep[e.g.,][]{JBarnes15}, planets in multiplanet systems \citep[e.g.,][]{Huber13}, and warm Jupiters \citep[e.g.,][]{Bourrier15}, but the sample sizes for each class are still too small to enable robust statistical investigations.

In this paper we present spin-orbit misalignment measurements for two hot Jupiters (HAT-P-41~b and WASP-79~b) and one warm Jupiter (Kepler-448~b) using Doppler tomography. When a planet transits a rotating star, the obscured stellar surface elements do not contribute to the formation of the rotationally broadened stellar line profile, resulting in a perturbation to the line profile. This is known as the Rossiter-McLaughlin effect \citep{Rossiter24,McLaughlin24}, and is typically interpreted as an anomalous radial velocity shift during the transit due to the changing line centroids \citep[e.g.,][]{Winn05}. In Doppler tomography, the star is sufficiently rapidly rotating, and the data obtained with sufficiently high spectral resolution, that the line profile perturbation can be spectroscopically resolved \citep[e.g.,][]{CollierCameron10,Johnson14}.

\section{Observations and Methodology}

\subsection{Observations}

Doppler tomographic analysis requires high signal-to-noise, high resolution spectra obtained with high time cadence during a transit. Our data on HAT-P-41~b and Kepler-448~b were obtained with the 9.2 m Hobby-Eberly Telescope (HET) at McDonald Observatory and its High-Resolution Spectrograph \citep[HRS;][]{Tull98}. These observations were obtained with a resolving power of $R=30,000$, and cover the range $\sim4770-6840$ \AA. Due to the HET's fixed altitude design it can typically only observe a given target for approximately one hour at a time, and so it is not possible to observe the entirety of a single transit. Instead, we observed portions of multiple transits (five of HAT-P-41~b and three of Kepler-448~b) and concatenated these datasets for the Doppler tomographic analysis. This is also the approach taken by \cite{Johnson14}. We obtained a total of 36 spectra (30 in transit) for HAT-P-41~b, and 30 (24 in transit) for Kepler-448~b. See Table~\ref{obstable} for more details of the observations.

The data on WASP-79~b were obtained by \cite{Addison13} using the 3.9 m Anglo-Australian Telescope (AAT) at Siding Spring Observatory, Australia. They used the CYCLOPS2 optical fiber bundle \citep{Horton12} and the University College London \'Echelle Spectrograph \citep[UCLES;][]{Diego90} to observe WASP-79 during the transit of 2012 December 23 UT. They obtained a total of 23 spectra, beginning just before ingress and continuing for about three hours after egress. These data have $R=70,000$ and wavelength coverage from 4550 \AA\,to 7350 \AA. See Table~\ref{obstable} and \cite{Addison13} for further details of the observations. 

\begin{deluxetable}{llccc}
\tabletypesize{\footnotesize}
\tablecolumns{5}
\tablewidth{0pt}
\tablecaption{Log of Observations \label{obstable}}
\tablehead{
\colhead{Planet} & \colhead{Date (UT)} & \colhead{Transit Phases} & \colhead{SNR} & \colhead{$N_{\mathrm{spec}}$} 
}

\startdata
Kepler-448 b & 2012 May 21 &  $0.66-0.85$ & 58 & 8 \\
WASP-79 b & 2012 Dec 23 &  $-0.06-1.54$ & 63 & 23 \\
Kepler-448 b & 2013 Mar 27 &  template & 17 & 6 \\
Kepler-448 b & 2013 Apr 25 &  $0.08-0.27$ & 43 & 8 \\
Kepler-448 b & 2013 May 13 &  $0.40-0.59$ & 57 & 8 \\
HAT-P-41 b & 2013 Jun 27 &  $0.09-0.35$ & 68 & 6 \\
HAT-P-41 b & 2013 Jul 08 &  $0.44-0.67$ & 68 & 6 \\
HAT-P-41 b & 2013 Jul 12 &  template & 59 & 6 \\
HAT-P-41 b & 2013 Jul 24 &  $0.04-0.25$ & 67 & 5 \\
HAT-P-41 b & 2013 Aug 04 &  $0.29-0.59$ & 62 & 7 \\
HAT-P-41 b & 2013 Aug 12 &  $0.51-0.77$ & 82 & 6 \\
\enddata

\tablecomments{Log of all Doppler tomographic observations. Observations of Kepler-448~b and HAT-P-41~b were obtained with the HET and HRS, and those of WASP-79~b with the AAT and UCLES. We define the ``transit phase'' such that it equals 0 at first contact and 1 at fourth contact, i.e. it is the fractional progress through the course of the transit. 
``Template'' denotes out-of-transit observations to fix the line shape. 
The quoted signal-to-noise ratio (SNR) is the mean SNR per pixel near 5500 \AA\,for all spectra taken on that night. $N_{\mathrm{spec}}$ is the number of spectra obtained during that night's observations.}

\end{deluxetable}

\subsection{Data Reduction and Analysis}

We reduced the HET data using standard IRAF tasks; this is described in more detail in \cite{Johnson14}. The AAT data were reduced using MATLAB routines as described in \cite{Addison13}.

Our methodology for the Doppler tomographic data preparation and analysis are the same as used in \cite{Johnson14} and \cite{Johnson15}. In summary, we first extract the average line profile from each spectrum using our own implementation of least squares deconvolution \citep{Donati97}; the extraction proceeds in a number of steps detailed in \cite{Johnson14}, utilizing a line mask with initial guesses for the line depths provided by a Vienna Atomic Line Database \citep{Ryabchikova15} spectral model based upon the literature stellar parameters. The final line depths are found by fitting our data, and are then used to extract the average line profiles. 

We model the time series line profiles by numerically integrating across the visible stellar disk, assuming a Gaussian line profile for each stellar surface element which is appropriately Doppler shifted assuming solid body rotation. The model accounts for the motion of the planetary disk during the finite exposures. See \cite{Johnson14} for full details on the generation of the model. Even though we ignore the effects of macroturbulence insofar as it is non-Gaussian, our model adequately reproduces the line shape even for moderately rapidly rotating stars like HAT-P-41 and WASP-79.

We generate posterior distributions for the transit parameters by exploring the likelihood space of model fits to the data using an affine-invariant Markov chain Monte Carlo \citep{GoodmanWeare10} as implemented in the Python package \texttt{emcee} \citep{ForemanMackey13}. We include the following parameters in the MCMC: spin-orbit misalignment $\lambda$, impact parameter $b$, projected stellar rotational velocity $v\sin i_{\star}$, orbital period $P$, transit epoch $T_0$, scaled semi-major axis $a/R_{\star}$, radius ratio $R_P/R_{\star}$, the width of the Gaussian intrinsic line profile of each stellar surface element, and two quadratic limb darkening coefficients. For the limb darkening coefficients we utilize the triangular sampling method of \cite{Kipping13}. We set Gaussian priors upon the limb darkening coefficients with the prior value found by interpolating $V$-band limb darkening values from \cite{Claret04} for ATLAS model atmospheres to the literature stellar parameters of each target star using the code \texttt{JKTLD}
 \citep{Southworth15}. We use the $V$ band as it approximates the region of the spectrum where we have both many stellar lines and high signal-to-noise in our spectra. Depending upon the state of prior knowledge of the system and the details of the Doppler tomographic dataset we also set Gaussian priors upon some of the other parameters, but as these vary from system to system we will note these in the sections describing each system. In cases where the literature source quotes an asymmetric uncertainty on a parameter, for simplicity we maintain a symmetric Gaussian prior, and conservatively set the prior width to the larger value of the literature uncertainty. Otherwise, we set uniform priors on the other parameters, with appropriate cut-offs to keep parameters in physically-allowed regions of parameter space: $P>0$, $0<R_P/R_{\star}<1$, $a/R_{\star}>0$, $|b|<1+R_P/R_{\star}$ (in order to ensure that a transit occurs), $v\sin i_{\star}>0$, and intrinsic line width $>0$. In all cases we ran MCMCs with 100 walkers for 100,000 steps each, and cut off the first 20,000 steps of burn-in, producing 8 million samples from the posterior distributions.

We will discuss our detailed procedures and results for each system in \S\ref{H41sec}, \S\ref{W79sec}, and \S\ref{Kep448sec} for HAT-P-41~b, WASP-79~b, and Kepler-448~b, respectively.
We note that for Kepler-448~b, we also simultaneously fit the {\it Kepler} transit photometry and literature radial velocity measurements along with the Doppler tomographic data; our methodology for this case is described in \S\ref{Kep448sec}. We do not perform such a full fit for either HAT-P-41~b or WASP-79~b due to the lack of availability of space-based photometry--or, indeed, any photometric observations beyond those presented in the discovery papers--for these targets.

\subsection{Error Analysis}

\begin{figure}
\epsscale{1.15}
\plotone{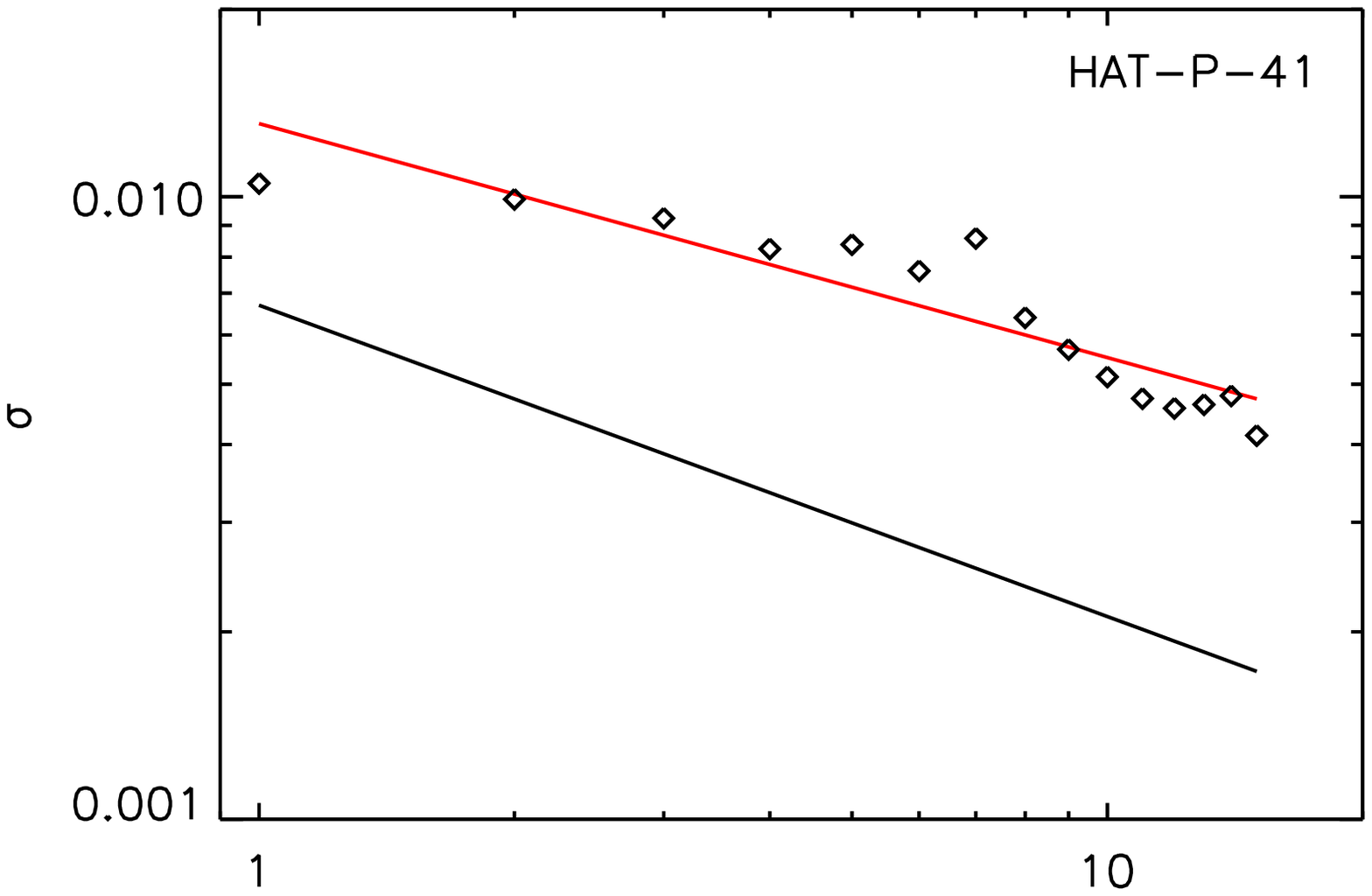}\\
\vspace{-24pt}
\plotone{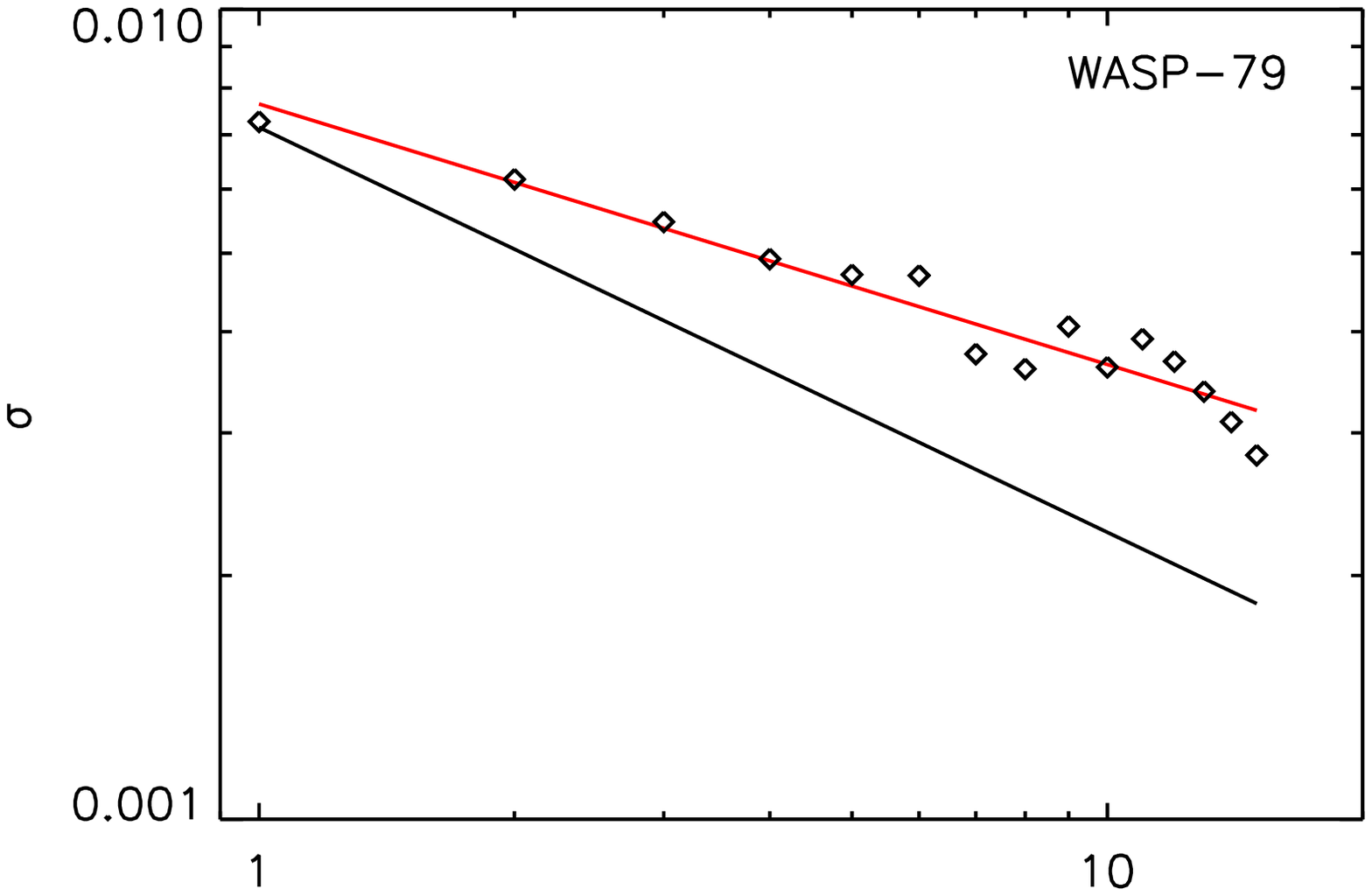}\\
\vspace{-24pt}
\plotone{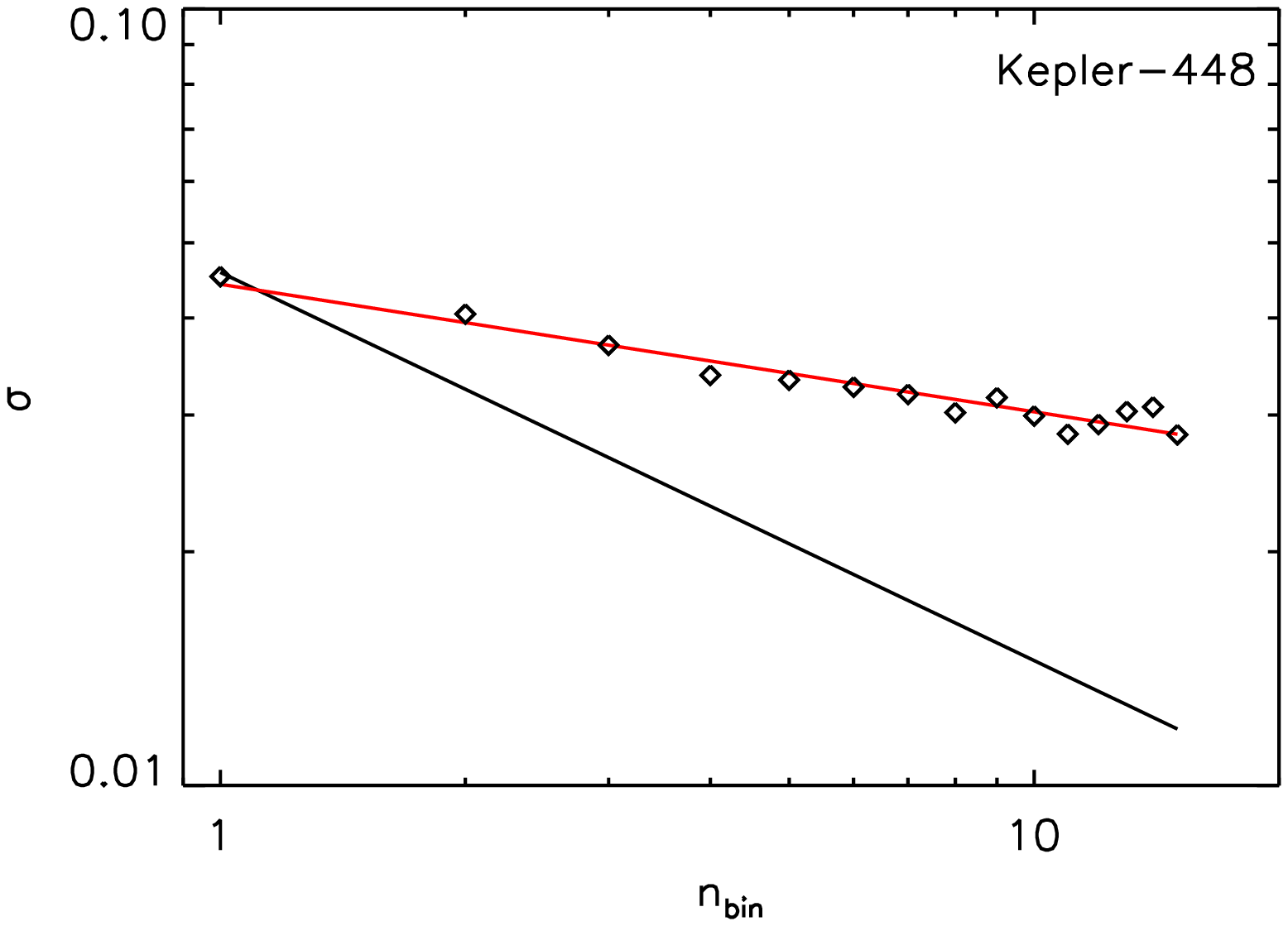}\\
\caption{Standard deviation of the continuum data after having $n_{\mathrm{bin}}$ pixels binned together (black diamonds), along with the expectation from the formal uncertainties on the data assuming white noise (black line) and a best-fit line to the black diamonds (red line). The panels show the results of this analysis for HAT-P-41 (top), WASP-79 (middle), and Kepler-448 (bottom).  \label{errorbinplots}} 
\end{figure}

In order to assess the presence of correlated noise in our Doppler tomographic data, we performed the following analysis. For each dataset, we chose the pixels with $|v|>v\sin i_{\star}+5$ km s$^{-1}$ (excluding the datasets from 2013 Mar 27 for Kepler-448, which is of much lower signal-to-noise than the other data, and for 2013 Jul 24 for HAT-P-41, which is contaminated by the solar spectrum; see \S\ref{H41sec}), and, for each spectrum, binned together $n_{\mathrm{bin}}=1, 2, \ldots 15$ pixels. For each bin size, we assessed the standard deviation $\sigma$ of the resulting binned time series line profile residuals. The results of this analysis are shown in Fig.~\ref{errorbinplots}. If the noise was strictly Gaussian, the standard deviation should decrease as the square root of the number of binned pixels, i.e., $\sigma\propto n_{\mathrm{bin}}^{-0.5}$; the black lines in Fig.~\ref{errorbinplots} depict $\sigma_{\mathrm{calc}}n_{\mathrm{bin}}^{-0.5}$, where $\sigma_{\mathrm{calc}}$ is the uncertainty on each data point calculated purely from photo-counting noise and the properties of the CCD, propagated through the line profile extraction process. Instead, for all three systems the data lie significantly above the expected line, indicating the presence of correlated noise. We fit straight lines to these data in log-log space (red lines in Fig.~\ref{errorbinplots}), and obtained that $\sigma\propto n_{\mathrm{bin}}^{-0.38}$ for HAT-P-41; $\sigma\propto n_{\mathrm{bin}}^{-0.32}$ for WASP-79; and $\sigma\propto n_{\mathrm{bin}}^{-0.16}$ for Kepler-448. In addition, it is apparent from Fig.~\ref{errorbinplots} that the analytic uncertainties well-reproduce the standard deviation of the continuum for WASP-79 and Kepler-448 (i.e., the black line intersects with the data at $n_{\mathrm{bin}}=1$), but not for HAT-P-41 b. This is likely due to the large systematics seen outside the line profile for this system (Fig.~\ref{hatp41dt}). Overall, this suggests that our calculated uncertainties are reasonable, but that there are correlations between pixels that are not captured by assuming independent Gaussian errors on each pixel. A similar analysis was performed by \cite{Bourrier15} for their Doppler tomographic data on Kepler-448 (see Fig.~5 of their work, but note that their cross-correlation functions are highly oversampled as compared to our line profiles), and have also been used for photometric transit observations \citep[e.g.,][]{Croll11}.

In order to account for the presence of correlated noise in our data, we use Gaussian process regression methodology, which was developed in part for this very purpose. Briefly, Gaussian process regression assumes some form for the covariance matrix, and several hyperparameters governing the amplitude and scale of the correlations are included in the MCMCs. See \cite{Gibson12} and \cite{Roberts13} for a more detailed introduction to Gaussian process regression. We implemented Gaussian process regression in our MCMCs using the \texttt{george} Python package\footnote{https://github.com/dfm/george} \citep{Ambikasaran14}. We used a two-dimensional Matern 3/2 kernel, which adds three parameters to our MCMCs: the amplitude of the Gaussian process $\alpha$, and characteristic length scales of the correlations along the velocity and time axes, $\tau_v$ and $\tau_t$, respectively. We set uniform priors on these parameters, with the requirement that all three be positive. We found that, when including the Gaussian process regression for fitting the time series line profiles, $v\sin i_{\star}$ was not well constrained, as the Gaussian process model was capable of reproducing the line profile regardless of the $v\sin i_{\star}$. We therefore simultaneously fit the time series line profiles using Gaussian process regression, and the average out-of-transit line profile without Gaussian process regression.

\section{HAT-P-41 \lowercase{b}}
\label{H41sec}

HAT-P-41~b is a hot Jupiter discovered by \cite{Hartman12}. It orbits a mildly rapidly rotating ($v\sin i_{\star}=19.6$ km s$^{-1}$) star every 2.69 days; the star has $T_{\mathrm{eff}}=6390$~K, and is thus above the 6250 K boundary where many planets have misaligned orbits \citep{Winn10}. The planet has a mass of $0.8\,M_J$ and, with a radius of $1.7\,R_J$, it is highly inflated. High-resolution imaging observations have identified two candidate stellar companions to HAT-P-41. The outer candidate, at $\sim4''$ \citep{Hartman12,Wollert15,WollertBrandner15,Evans16,Ngo16}, has a proper motion inconsistent with either a bound companion or a background object; it may therefore be a foreground object with its own significant proper motion \citep{Evans16}. An inner candidate companion was found at $\sim1''$ by \cite{Evans16}, but with only a single observation there is no proper motion information to assess whether this object is bound to HAT-P-41 or not. Although this object falls within the $2''$ HRS fibers, at $\Delta r=4.42$ it is too faint to contribute significantly to the flux or dilute the transit signal, and so we neglect it in our analysis. There are no published long-term radial velocities for HAT-P-41 that could enable a search for trends due to long-period planetary companions. We list the relevant parameters of the system in Table~\ref{H41table}. The $v\sin i_{\star}$ value is high enough for the rotationally broadened line profile to be spectroscopically resolved--i.e., the full width of the line profile is $\gtrsim3-4$ resolution elements--and so this system is amenable to Doppler tomographic observations.

\begin{figure}
\epsscale{1.15}
\plotone{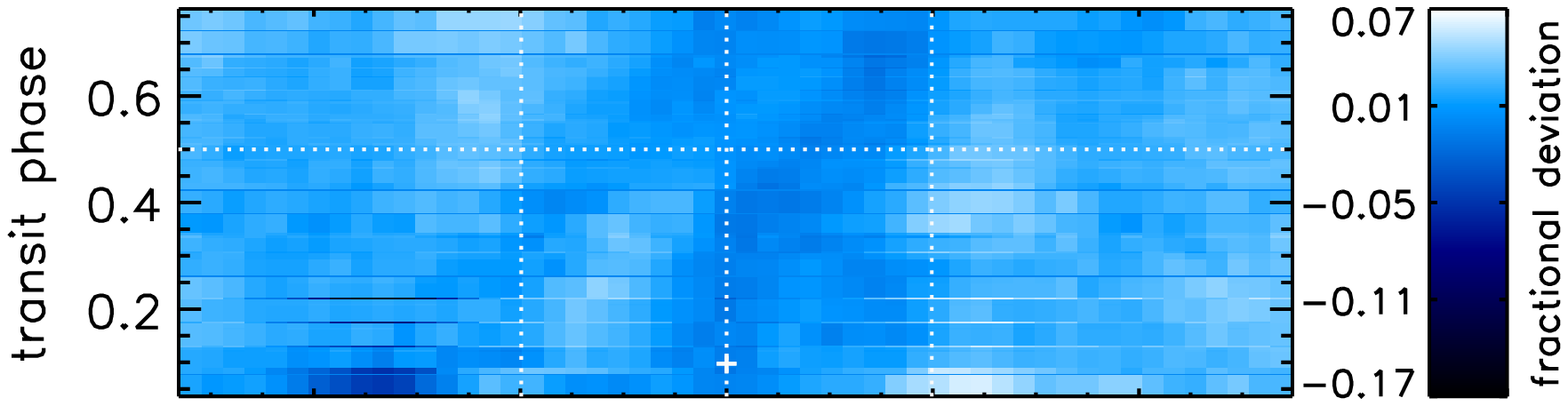}
\plotone{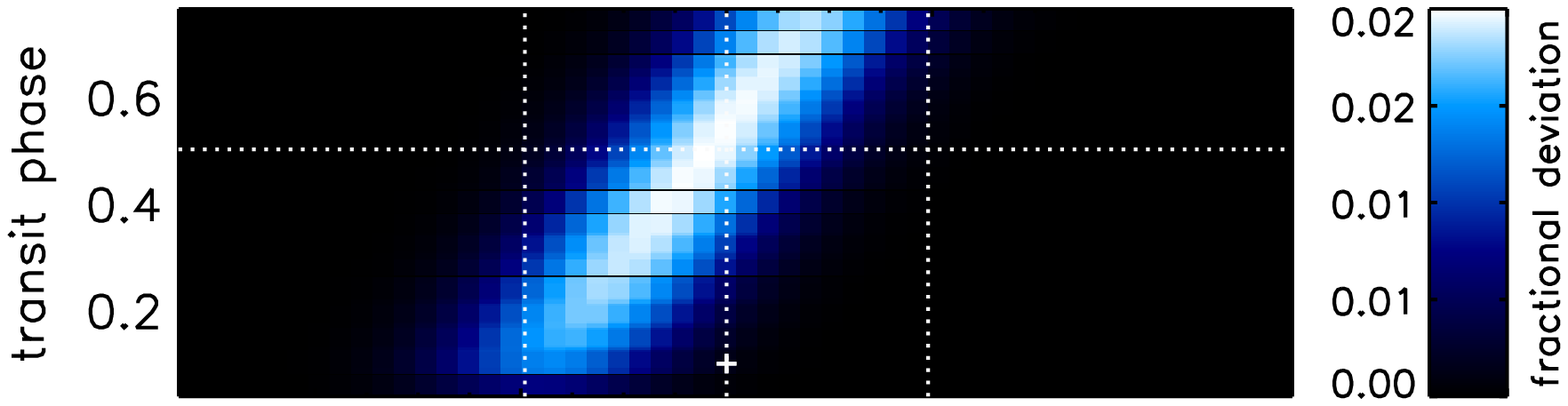}
\plotone{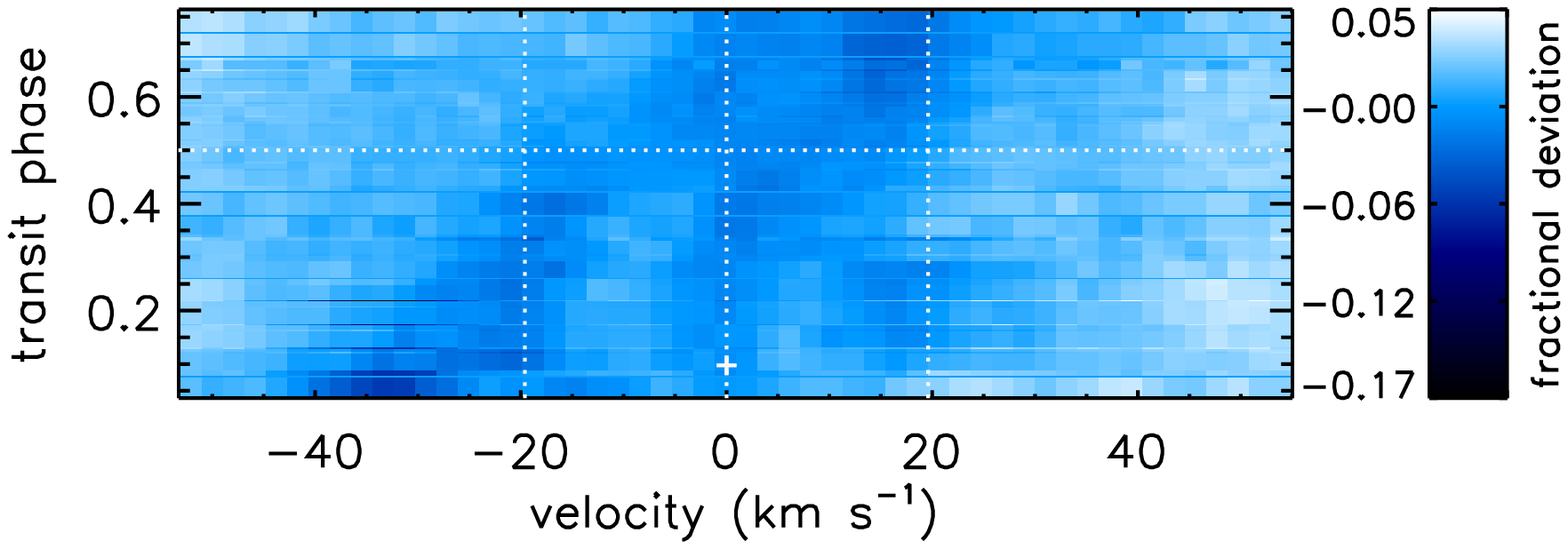}

\caption{Time series line profile residuals for HAT-P-41~b. In all panels, time increases from bottom to top, and each color scale row shows the deviation in the line profile at that time from the model stellar line profile. Bright regions denote shallower regions of the line, and so the line profile perturbation due to the transit manifests as a bright streak. Vertical dashed lines mark the center of the line profile ($v=0$) and the edges of the line profile at $v=\pm v\sin i_{\star}$, a horizontal dashed line marks the time of mid-transit, and the small crosses mark first through fourth contacts (although only second contact is included in the time span of these observations). We define the ``transit phase'' such that it equals 0 at first contact and 1 at fourth contact, and the ``fractional deviation'' is the fractional difference between each line profile and the model line profile. Top: data from all five HET-observed partial transits combined. The transit signature is the bright streak running from lower left to upper right, indicating a prograde orbit. Middle: model corresponding to our best-fit solution from the analysis of the shifted and binned time series line profile residuals (Fig.~\ref{hatp41shift}). Bottom: residuals after the best-fit model has been subtracted. The residuals are not completely flat, potentially indicating that systematics could bias our solution.   \label{hatp41dt}} 

\end{figure}

Parts of five transits were observed with the HET and HRS between 2013 June and August. We summarize the observations in Table~\ref{obstable}, and show the time series line profile residuals--i.e., the deviation of the time series line profiles from the out-of-transit line profile--in the top left panel of Fig.~\ref{hatp41dt}. It is apparent by inspection that the orbit of HAT-P-41~b is prograde, as the line profile perturbation begins the transit over the blueshifted wing of the line and moves redward over the course of the transit.

We do not, however, obviously detect the line profile perturbation in all of the datasets; indeed, it is most strongly detected only in the datasets from 2013 June 27 and July 24 UT, which together cover approximately the first one-third of the transit. We have been unable to determine the reason for this. These datasets have similar signal-to-noise to those from June 27 and July 24, and there is nothing else obvious to mark them as different. After extensive investigation and modifications to our line profile extraction procedure we have been unable to positively identify the source of this issue (or even whether it is astrophysical or instrumental), or to correct it. 

A further complication is that there is an additional line profile component due to scattered moonlight in the data obtained on 2013 July 24. It, however, does not significantly overlap with the stellar line profile; it is centered at $\sim-35$ km s$^{-1}$ (in the stellar barycentric rest frame), while the star has $v\sin i_{\star}=19.6 \pm 0.5$ km s$^{-1}$ \citep{Hartman12}, and so we can safely neglect this contaminating line profile.

Even when fitting only the data from 2013 June 27 and July 24 using Gaussian process regression methodology, and setting a prior upon $b$ using the value found by \cite{Hartman12}, our fitting code would not converge. Instead, we used alternate methodology to measure the spin-orbit misalignment of HAT-P-41~b. In \cite{Johnson14}, we introduced a method to optimally bin together Doppler tomographic data. In short, we take advantage of the fact that, in the absence of differential rotation and assuming a circular orbit \citep[which is the case for HAT-P-41~b:][]{Hartman12}, the speed of the line profile perturbation across the line profile is a constant--i.e., in a plot displaying the time series line profile residuals (e.g., Fig.~\ref{hatp41dt}), the path of the line profile perturbation is a straight line. If we shift the line profile residuals from each spectrum in velocity space such that the line profile perturbation lies at the same velocity in every spectrum, and bin together all of the spectra, we will obtain a higher signal-to-noise detection of the line profile perturbation. If, on the other hand, we choose an incorrect value for the slope of this path, the line profile perturbations from the different spectra will tend to average out, leaving us with no signal.

\begin{figure}
\epsscale{1.15}
\plotone{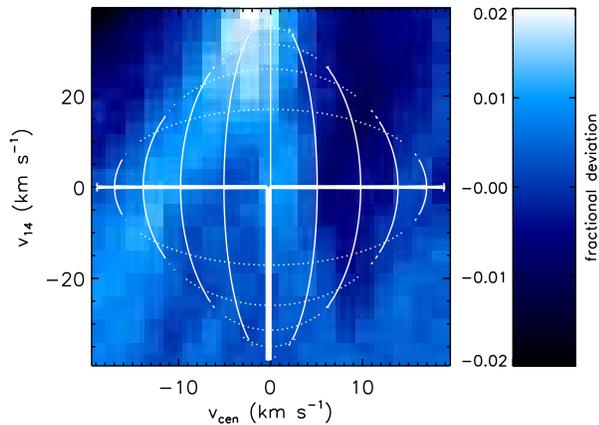}

\caption{Shifted and binned time series line profile residuals for HAT-P-41~b, as described in the text. $v_{\mathrm{cen}}$ is the velocity of the line profile perturbation at mid-transit, and $v_{14}$ is the difference between the velocity of the line profile perturbation at egress and ingress. The contours show the mapping between these variables and $\lambda$ and $b$. Solid contours show, from left to right, values of $\lambda = -75^{\circ}$, $-60^{\circ}$, $-45^{\circ}$, $-30^{\circ}$, $-15^{\circ}$, $0^{\circ}$, $15^{\circ}$, $30^{\circ}$, $45^{\circ}$, $60^{\circ}$, and $75^{\circ}$ in the upper half of the plot, and $\lambda = -105^{\circ}$, $-120^{\circ}$, $-135^{\circ}$, $-150^{\circ}$, $-165^{\circ}$, $\pm180^{\circ}$, $165^{\circ}$, $150^{\circ}$, $135^{\circ}$, $120^{\circ}$, and $105^{\circ}$ in the bottom half of the plot. Dotted contours show, moving outward (up and down) from the center line of the plot, $b = 0.9$, 0.75, 0.6, 0.45, and 0.3. Note that the relationship between $v_{\mathrm{cen}}$ , $v_{14}$ and $\lambda$, $b$ is double-valued; only the solution appropriate to HAT-P-41~b is shown here. The transit signature is the bright splotch near the top, indicating a prograde orbit.   \label{hatp41shift}} 

\end{figure}

In order to measure the spin-orbit misalignment of HAT-P-41 b, we perform this operation for all physically allowed values of the slope of this path, parameterized as $v_{14}$, the difference in velocity between the locations of the line profile perturbation at egress and ingress($|v_{14}|<2v\sin i_{\star}$); positive values of $v_{14}$ correspond to prograde orbits, negative values to retrograde orbits. The new velocity to which all of the spectra are shifted is parameterized as $v_{\mathrm{cen}}$ , the velocity of the line profile perturbation at mid-transit. We show these shifted and binned time series line profile residuals in Fig.~\ref{hatp41shift}. For this analysis we used all five nights of in-transit data, and the fact that we see a relatively compact peak, rather than a long streak, indicates that we do indeed have the line profile perturbation signal in all five datasets.

We extracted transit parameters from the shifted and binned data using the same methodology as was used by \cite{Johnson14} for similarly-treated data on Kepler-13 Ab. We used a custom-built MCMC (not \texttt{emcee}) to fit a model of the line profile perturbation to the data; the model was computed using the same methods as described earlier, and shifted and binned in the same manner as the data. In order to make the MCMC converge, we had to fix $b$, $v\sin i_{\star}$, and $R_P/R_{\star}$ to the values found by \cite{Hartman12}. The MCMC used the following jump parameters: $\lambda$, $P$, the transit duration $\tau_{14}$ and epoch, and two quadratic limb darkening coefficients, treated as described earlier. We set Gaussian priors upon all of the parameters except for $\lambda$, with values and widths taken from \cite{Hartman12}. We ran four chains each for 150,000 steps, cutting off the first 20,000 steps of burn-in.

\begin{deluxetable*}{lccc}
\tabletypesize{\footnotesize}
\tablecolumns{4}
\tablewidth{0pt}
\tablecaption{System Parameters of HAT-P-41 \label{H41table}}
\tablehead{
\colhead{Parameter} & \colhead{Value} & \colhead{Prior} & \colhead{Source} 
}

\startdata
Stellar Parameters & & & \\
$T_{\mathrm{eff}}$ (K) & $6390 \pm 100$ & \ldots & \cite{Hartman12} \\
$M_{\star}\,(M_{\odot})$ & $1.418 \pm 0.047$ & \ldots & \cite{Hartman12}  \\
$R_{\star}\,(R_{\odot})$ & $1.683_{-0.036}^{+0.058}$ & \ldots & \cite{Hartman12}  \\
\hline
Planetary Parameters & &  \\
$M_P\,(M_J)$ & $0.80 \pm 0.10$ & \ldots & \cite{Hartman12}  \\
$R_P\,(R_J)$ & $1.685_{-0.051}^{+0.076}$ & \ldots & \cite{Hartman12}  \\
\hline
MCMC Inputs & &  \\
$P$ (days) & $2.694047 \pm 0.000004$ & $\mathcal{G}(2.694047,0.000004)$ & \cite{Hartman12}  \\
$T_0$ (BJD) & $2454983.8617 \pm 0.0011$ & $\mathcal{G}(2454983.8617,0.0011)$ & \cite{Hartman12}  \\
$R_P/R_{\star}$ & $0.1028 \pm 0.0016$ & $\mathcal{F}(0.1028)$ & \cite{Hartman12}  \\
$\tau_{14}$ (d) & $0.1704 \pm 0.0012$ & $\mathcal{G}(0.1704,0.0012)$ & \cite{Hartman12} \\
$b$ & $0.222_{-0.093}^{+0.088}$ & $\mathcal{F}(0.222)$  & \cite{Hartman12}  \\
$v\sin i_{\star}$ (km s$^{-1}$) & $19.6 \pm 0.5$ & $\mathcal{F}(19.6)$ & \cite{Hartman12} \\
\hline
Measured Parameters & &  \\
$\lambda\,(^{\circ})$ & $-22.1_{-6.0}^{+0.8}$ & $\mathcal{U}(-180,180)$ & this work  \\
\enddata

\tablecomments{Uncertainties are purely statistical and do not take into account systematic sources of error. The parameters in the MCMC Inputs section are the MCMC parameters where we incorporated prior knowledge; Measured Parameters are those that we measured directly with the MCMC. The Prior column lists the type of prior used for each parameter in the MCMC. Notation: $\mathcal{U}(x,y)$: uniform prior between $x$ and $y$. $\mathcal{G}(x,y)$: Gaussian prior with mean $x$ and standard deviation $y$. $\mathcal{F}(x)$: value fixed to $x$. Parameters with no prior type listed were not used in our MCMC and are quoted here for informational purposes only.}

\end{deluxetable*}

We summarize our results in the bottom section of Table~\ref{H41table}. We find a value of the spin-orbit misalignment of $\lambda = -22.1_{-2.4}^{+0.3 \circ}$. We emphasize that the uncertainties on this value may be underestimated, as we did not use Gaussian process regression for this analysis; furthermore, as can be seen in the bottom panel of Fig.~\ref{hatp41dt}, significant systematics remain after the subtraction of the best-fit model, suggesting that our result could be biased. \cite{Johnson14} investigated the effects of uncharacterized macroturbulence and differential rotation on Doppler tomographic data, and found that these effects tended to have systematic effects on $\lambda$ a factor of two larger than the formal uncertainties. In order to be conservative in the presence of these factors and possible other biases we therefore inflate our uncertainties by a factor of 2.5 and adopt $\lambda = -22.1_{-6.0}^{+0.8 \circ}$ as the best value that can be obtained from the current data. The orbit of HAT-P-41 b is thus somewhat misaligned, as is typical for planets around stars above the Kraft break \citep[HAT-P-41 has $T_{\mathrm{eff}} = 6390$~K;][]{Hartman12}. 

\section{WASP-79 \lowercase{b}}
\label{W79sec}

WASP-79~b is a hot Jupiter that was discovered by \cite{Smalley12}. It orbits a relatively bright ($V=10.1$) F5 star with a period of 3.662 days. The star is mildly rapidly rotating, with $v\sin i_{\star}=19.1 \pm 0.7$ km s$^{-1}$. \cite{Smalley12} produced two sets of system parameters, one assuming a main sequence primary (i.e., enforcing the main sequence $M_{\star}-R_{\star}$ relation in their global fit) and one assuming a non-main sequence host (no $M_{\star}-R_{\star}$ relation assumed). There are no additional known objects in the WASP-79 system; high angular resolution lucky imaging observations found no candidate stellar companions \citep{Evans16}, but there are no published long-term radial velocity observations to check for outer planetary companions. We quote the literature parameters of the WASP-79 system in Table~\ref{W79table}.

WASP-79 was observed using radial velocity Rossiter-McLaughlin methodology by \cite{Addison13}, who measured a spin-orbit misalignment of $\lambda=-106_{-13}^{+19 \circ}$ \citep[and a second solution of $\lambda=-84_{-30}^{+23 \circ}$ by assuming the non-main sequence parameters from][]{Smalley12}. Due to the relatively rapid stellar rotation and the fact that the high wavelength stability necessary for precise radial velocity observations was obtained using simultaneous ThXe calibration (rather than through the use of an iodine cell), these data are also amenable to Doppler tomographic analysis. Here we reanalyze these data using Doppler tomographic methodology.

\begin{figure}
\epsscale{1.15}
\plotone{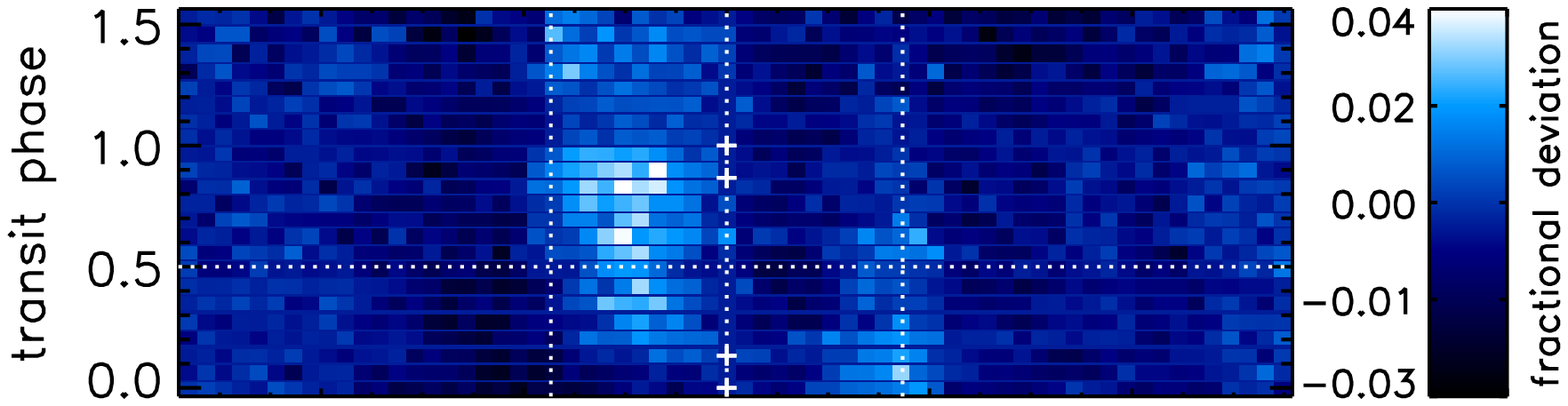}\\
\plotone{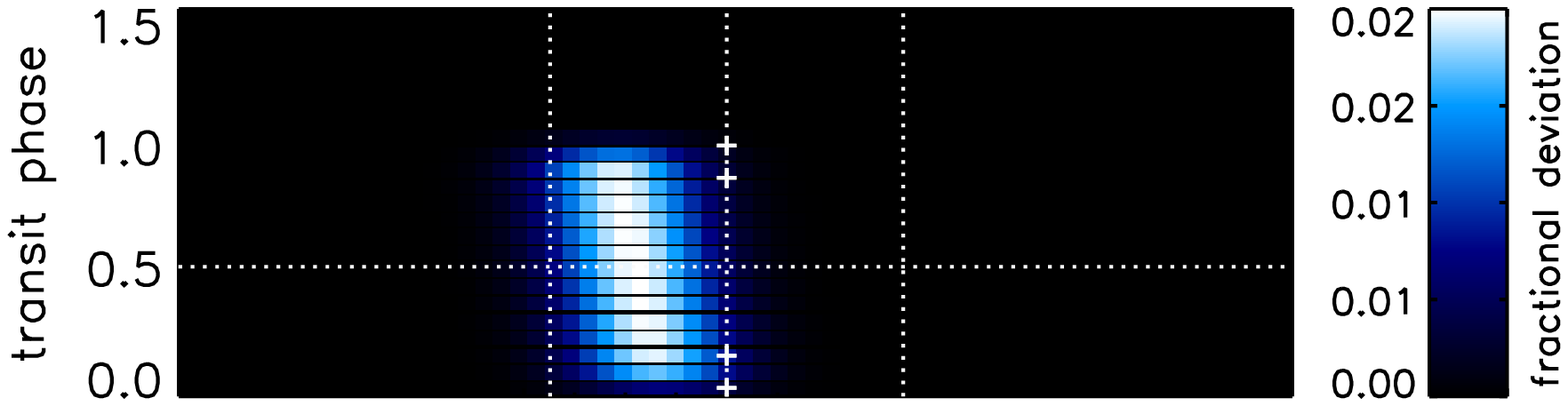}\\
\plotone{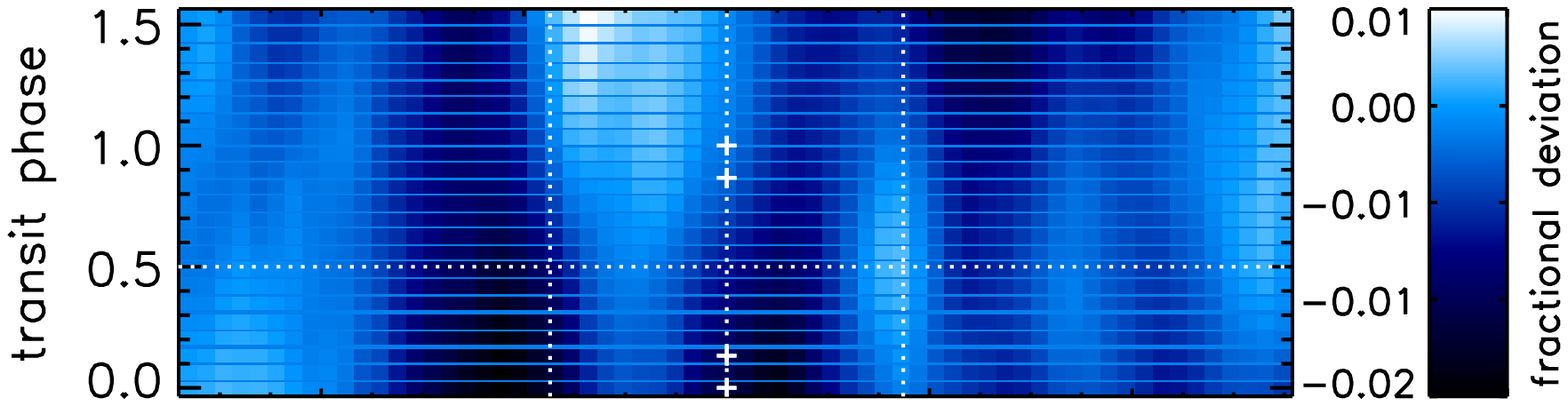}\\
\plotone{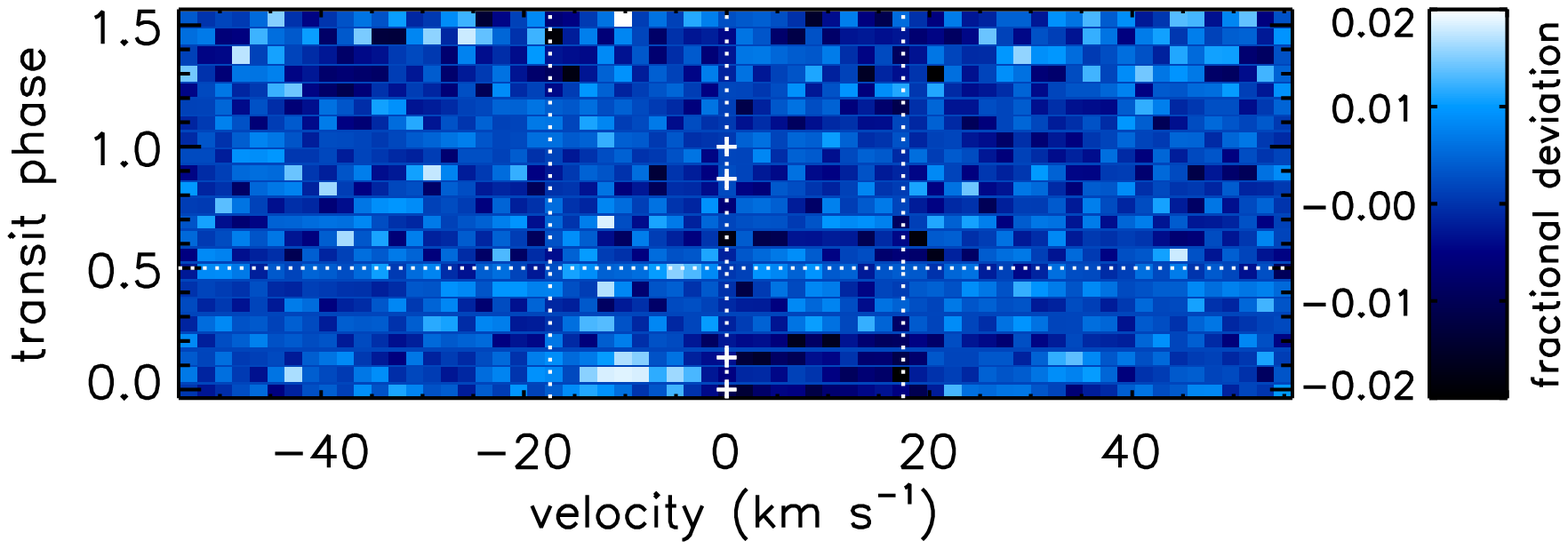}\\
\caption{Time series line profile residuals showing the 2012 December 23 transit of WASP-79~b. Top: the data, showing the line profile perturbation due to the transit (the bright vertical streak to the left of center). Top middle: best-fit model of the line profile perturbation. Bottom middle: best-fit Gaussian process regression model. Bottom: residuals after the best-fit transit and Gaussian process regression models have been subtracted off from the data. Sources of the systematics in these data are discussed in the text. Notation on this plot is the same as on Fig.~\ref{hatp41dt}. \label{wasp79dt}} 
\end{figure}

We show the time series line profile residuals in Fig.~\ref{wasp79dt}; we easily detect the transit. The path of the line profile perturbation across the time series line profile residuals is nearly vertical, indicating a highly-inclined orbit as found by \cite{Addison13}. As can be seen in Fig.~\ref{wasp79dt}, there are also some residual systematics in the time series line profile residuals. Two types of systematics are visible: time-invariant--i.e., bright streaks running the full length of the figure--and time-dependent--i.e., streaks or blobs that change or disappear over time. The former is most easily seen after the transit has ended, when the blue-shifted half of the line profile appears brighter than the red-shifted half; this is caused by a systematic mismatch between the shape of the model line profile and the observed line profile. The latter can be seen in the bright streak near +20 km s$^{-1}$, beginning at the start of the dataset and dwindling away by mid-transit. A corresponding dark streak at -20 km s$^{-1}$ is most visible in the residuals (bottom panel of Fig.~\ref{wasp79dt}). These residuals are caused by a time-dependent velocity mismatch between the model and observed line profiles, resulting in a positive residual on one side of the line profile, and a negative residual on the other. As can also be seen in Fig.~\ref{wasp79dt}, these systematics are mostly accounted for by the Gaussian process regression model.

\floattable
\rotate
\begin{deluxetable*}{lccccc}
\tabletypesize{\footnotesize}
\tablecolumns{6}
\tablewidth{0pt}
\tablecaption{System Parameters of WASP-79 \label{W79table}}
\tablehead{
\colhead{Parameter} & \colhead{Value} & \colhead{Prior} & \colhead{Value} & \colhead{Prior} & \colhead{Source}  \\
\colhead{} & \colhead{(MS; preferred)} & \colhead{(MS)} & \colhead{(non-MS)} & \colhead{(non-MS)} & \colhead{} 
}

\startdata
Stellar Parameters & & & & \\
$T_{\mathrm{eff}}$ (K) & $6600 \pm 100$ &  \ldots & $6600 \pm 100$ & \ldots & \cite{Smalley12}  \\
$M_{\star}\,(M_{\odot})$ & $1.56 \pm 0.09$ & \ldots & $1.52 \pm 0.07$ & \ldots & \cite{Smalley12}  \\
$R_{\star}\,(R_{\odot})$ & $1.64 \pm 0.08$ & \ldots & $1.91 \pm 0.09$ & \ldots & \cite{Smalley12}  \\
$v\sin i_{\star}$ (km s$^{-1}$) & $19.1 \pm 0.7$ & $\mathcal{U}(0,\infty)$ & $19.1 \pm 0.7$ & $\mathcal{U}(0,\infty)$ & \cite{Smalley12} \\
\hline
Planetary Parameters & & & & \\
$M_P\,(M_J)$ & $0.90 \pm 0.09$ & \ldots & $0.90 \pm 0.08$ & \ldots & \cite{Smalley12}  \\
$R_P\,(R_J)$ & $1.70 \pm 0.11$ & \ldots & $2.09 \pm 0.14$ & \ldots & \cite{Smalley12}  \\
\hline
MCMC Inputs & & & & \\
$P$ (days) & $3.6623817 \pm 0.0000051$ & $\mathcal{G}(3.6623817,0.0000051)$ & $3.6623866 \pm 0.0000085$ & $\mathcal{G}(3.6623866,0.0000085)$ & \cite{Smalley12}  \\
$T_0$ (BJD\_TDB-2450000)\tablenotemark{a} & $2455545.2356 \pm 0.0013$ & $\mathcal{G}(2455545.2356,0.0013)$ & $2455545.2361 \pm 0.0015$ & $\mathcal{G}(2455545.2361,0.0015)$ & \cite{Smalley12}  \\
$R_P/R_{\star}$\tablenotemark{a} & $0.1071 \pm 0.0024$ & $\mathcal{G}(0.1071,0.0024)$ & $0.1126 \pm 0.0028$ & $\mathcal{G}(0.1126,0.0028)$ & \cite{Smalley12}  \\
$a/R_{\star}$\tablenotemark{a} & $7.1 \pm 1.1$ & $\mathcal{G}(7.1,1.1)$ & $6.05 \pm 0.52$ & $\mathcal{G}(6.05,0.52)$  & \cite{Smalley12}  \\
$b$ & $0.570 \pm 0.052$ & $\mathcal{U}(-(1+R_P/R_{\star}),1+R_P/R_{\star})$ & $0.706\pm  0.031$ & $\mathcal{U}(-(1+R_P/R_{\star}),1+R_P/R_{\star})$ & \cite{Smalley12}  \\
\hline
\multicolumn{1}{l}{Rossiter-McLaughlin Parameter}  \\
$\lambda$ ($^{\circ}$) & $-106^{+19}_{-13}$ & $\mathcal{U}(-180,180)$ & $-84_{-30}^{+23}$ & $\mathcal{U}(-180,180)$ & \cite{Addison13} \\
$v\sin i_{\star}$ (km s$^{-1}$) & $17.5_{-3.0}^{+3.1}$ & $\mathcal{U}(0,\infty)$ & $16.0 \pm 3.7$ & $\mathcal{U}(0,\infty)$  & \cite{Addison13}  \\
\hline
Measured Parameters & & & & \\
$b$ & $0.538 \pm 0.047$ & $\mathcal{U}(-(1+R_P/R_{\star}),1+R_P/R_{\star})$ & $0.571_{-0.05}^{+0.07}$ & $\mathcal{U}(-(1+R_P/R_{\star}),1+R_P/R_{\star})$ & this work  \\
$\lambda$ ($^{\circ}$) & $-99.1_{-3.9}^{+4.1}$ & $\mathcal{U}(-180,180)$ & $-96.8_{-4.1}^{+5.7}$ & $\mathcal{U}(-180,180)$ & this work  \\
$v\sin i_{\star}$ (km s$^{-1}$) & $17.41_{-0.12}^{+0.20}$ & $\mathcal{U}(0,\infty)$ & $17.45_{-0.15}^{+0.35}$ & $\mathcal{U}(0,\infty)$ & this work  \\
intrinsic line width (km s$^{-1}$) & $5.35_{-0.19}^{+0.18}$ & $\mathcal{U}(0,\infty)$ & $5.31_{-0.22}^{+0.23}$ & $\mathcal{U}(0,\infty)$ & this work \\
\enddata

\tablecomments{The parameters in the MCMC Inputs section are the MCMC parameters where we incorporated prior knowledge; Measured Parameters are those that we measured directly with the MCMC. Rossiter-McLaughlin Parameters are those measured by \cite{Addison13} in their analysis. The Prior column lists the type of prior used for each parameter in the MCMC. Notation: $\mathcal{U}(x,y)$: uniform prior between $x$ and $y$. $\mathcal{G}(x,y)$: Gaussian prior with mean $x$ and standard deviation $y$. Parameters with no prior type listed were not used in our MCMC and are quoted here for informational purposes only. As discussed in the text, we performed separate MCMC fits using the main sequence (MS) and non-main sequence (non-MS) system parameters from \cite{Smalley12}, but argue that the main sequence solution is likely to be the correct solution.}
\tablenotetext{a}{Calculated analytically from the parameters given in the literature source.}

\end{deluxetable*}

The two solutions for the system parameters from \cite{Smalley12} have different values of several of our MCMC parameters, most notably $b$ ($0.570 \pm 0.052$ for the main sequence case and $0.706 \pm 0.031$ for the non-main sequence solution), but also $R_P/R_{\star}$ and $a/R_{\star}$. We consequently ran two separate MCMCs, one assuming the main sequence values from \cite{Smalley12} as the starting values and priors on $P$, $T_0$, $R_P/R_{\star}$, and $a/R_{\star}$, and the other using the non-main sequence values. Due to the highly-inclined orbit of WASP-79~b, the path of the line profile perturbation across the line profile is highly sensitive to the impact parameter $b$, and so we set a uniform prior upon this parameter. 

We list the parameters that we found for WASP-79~b in the bottom section of Table~\ref{W79table}. The values of $\lambda$ that we measured are consistent with those found by \cite{Addison13}, but a factor of a few more precise (assuming the Gaussian process regression allows an accurate estimate of the uncertainties). 

Our results are insensitive to the choice of priors \citep[i.e., the main sequence vs. non-main sequence solutions of][]{Smalley12}; our best-fit values of $\lambda$, $b$, $v\sin i_{\star}$, and the intrinsic Gaussian line width all vary by $<0.5\sigma$ between the two solutions. Interestingly, the value of $b$ that we found for the main sequence priors ($b=0.538 \pm 0.047$) is consistent ($0.5\sigma$ difference) with that found by \cite{Smalley12}, but that for the non-main sequence priors ($b=0.571_{-0.05}^{+0.07}$) is somewhat ($1.8\sigma$) discrepant with that from \cite{Smalley12}, but in agreement with our value of $b$ from the main sequence priors. Since our measurement of $b$ is largely independent of the choice of priors, and both cases agree with the main sequence solution of \cite{Smalley12}, we conclude that this is likely to be the correct solution for the system and adopt the solution with main sequence priors as the preferred solution.

\cite{Brown16} also recently presented Doppler tomographic observations of WASP-79; they found $\lambda=-95.2_{-1.0}^{+0.9 \circ}$, a value that compatible ($0.9\sigma$ difference) with ours. They also found $b=0.50 \pm 0.02$, again compatible with our value.

\section{Kepler-448 \lowercase{b}}
\label{Kep448sec}

Kepler-448~b (aka KOI-12.01) is a warm Jupiter discovered by \emph{Kepler} \citep{Borucki10}; it was first identified as a planet candidate by \cite{Borucki11}. It was subsequently validated by \cite{Bourrier15}, using their own Doppler tomographic observations with the SOPHIE spectrograph \citep{Perruchot08} on the 1.93 m telescope at the Observatoire de Haute-Provence, France. It has an orbital period of 17.9 days and, with a host star magnitude of $Kp=11.353$, it is one of the brightest stars known to host a transiting warm Jupiter that is not on a highly eccentric orbit. Additionally, with $v\sin i_{\star}=60$ km s$^{-1}$ \citep{Bourrier15}, it is a good target for Doppler tomography.

We obtained our own Doppler tomographic dataset on this system using the HET and HRS. These data span parts of three transits, on 2012 May 21, 2013 April 25, and 2013 May 13 UT, as well as an out-of-transit template observation on 2013 March 27 UT. Again, due to the fixed-altitude design of the HET, we could only observe small parts of the $\sim7$-hour-long transit at once. See Table \ref{obstable} for more details of the observations. 

A complication for the Doppler tomographic analysis was that two of the four datasets, the in-transit data from 2013 April 25 and the template data, 
were contaminated with the solar spectrum reflecting off of the Moon. 
This resulted in a narrow absorption line profile superposed upon the rotationally broadened line profile of Kepler-448; unlike for HAT-P-41, the contaminating solar line profile lies within the line profile of Kepler-448. We dealt with this complication by including this additional line profile in our model. We assumed that the solar line profile was unresolved and thus we could model it as identical to the instrumental line profile. We added it to the model line profiles of Kepler-448~b on these nights, and added two additional parameters with uniform priors to the MCMC to govern the behavior of this line: the line depth and central velocity.

In order to fully characterize the system, we performed a joint fit to the Doppler tomographic data, {\it Kepler} light curve, and literature radial velocity measurements. Kepler-448 was observed by the {\it Kepler} spacecraft for its entire prime mission (Quarters 0 through 17). It was observed with short cadence (1 minute integrations) photometry for every quarter except for Quarter 1, when it was observed in long cadence mode (30 minute integrations).
We obtained the {\it Kepler} light curve for Kepler-448 from the MAST archive\footnote{http://archive.stsci.edu/kepler/}, and used the \texttt{PyKE} \citep{StillBarclay12} software tool \texttt{kepcotrend} to remove systematic trends in the data using the cotrending basis vectors provided by the Kepler team\footnote{https://archive.stsci.edu/kepler/cbv.html}. We then divided each flux value by the mean flux in that quarter to produce normalized light curves for each quarter, and spliced these together to produce a full light curve spanning more than four years of observations.

\begin{figure}
\epsscale{1.25}
\plotone{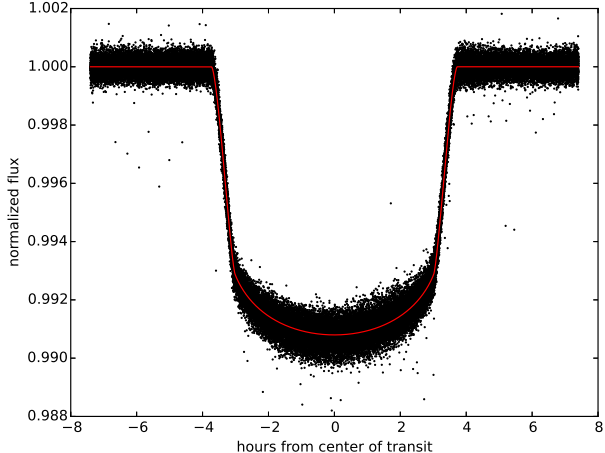}
\caption{\emph{Kepler} short-cadence data in and near the transit of Kepler-448~b, phased to the orbital period and with our best-fit photometric transit model overplotted in red. We removed the stellar variability as described in the text, and excluded two transits for which this process resulted in highly distorted transit shapes; this figure includes data from the other 77 transits observed by \emph{Kepler} in short cadence.  \label{koi12trlc}} 
\end{figure}

The light curve of Kepler-448 shows rotational modulation with an amplitude of $\sim0.1-0.2\%$ and a period of $\sim1.5$ days. We investigate the properties of this modulation in \S\ref{rotationsection} below, but in order to model the photometric transit data it was necessary to remove this variability. For each transit we fit a quadratic function to the \emph{Kepler} short cadence out-of-transit flux within one transit duration (7.4 hours) of the transit center, and divided out this fit to produce a flattened, normalized transit light curve. This produced good results for most of the transits, except for two where the shape was highly distorted due to lack of data before and/or after the transit, which we excluded from the dataset used in our fits. The remaining transits often still show some low-level distortions, but these should average out when fitting many transits. Indeed, our final best-fit transit model (Fig.~\ref{koi12trlc}) well reproduces the data. 
The photometric fitting routine is derived from that used in \cite{Mann16a} and \cite{Mann16b}, using model light curves generated using the \texttt{batman} package \citep{Kreidberg15}, now coupled to the Doppler tomographic and radial velocity fitting.

\begin{figure}
\epsscale{1.25}
\plotone{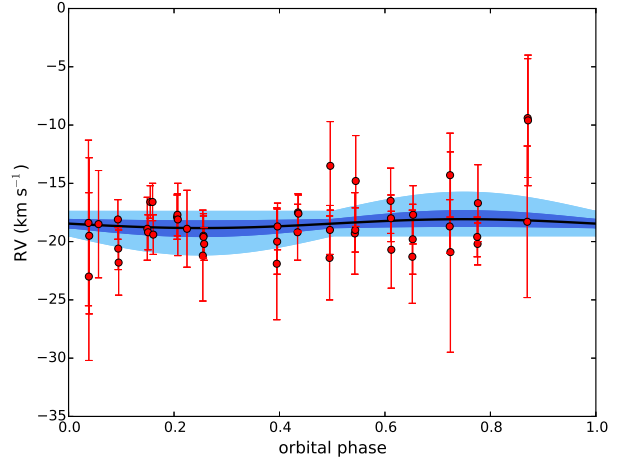}
\caption{Radial velocity measurements of Kepler-448 from \cite{LilloBox15}, phased on the transit period and with the best-fit model from our MCMC overplotted in black. The dark and light blue regions show the $1\sigma$ and $3\sigma$ credible regions, respectively, due to uncertainty in $K$ and $\gamma$; we neglect uncertainties in other parameters (e.g., $P$) because they are proportionally much smaller. \label{koi12rvfig}} 
\end{figure}

We also included the radial velocity observations of \cite{LilloBox15} in our fit. They obtained 47 radial velocity measurements of Kepler-448 over a span of 114 days using the CAFE spectrograph \citep{Aceituno13} on the 2.2 m telescope at the Calar Alto Observatory, Spain. They obtained an upper limit on the mass of Kepler-448~b of $25.2 \pm 3.7\,M_J$, limiting it to be a planet or brown dwarf \citep[using a smaller dataset,][obtained a $3\sigma$ limit on the mass of $<8.7\,M_J$]{Bourrier15}. We show the data from \cite{LilloBox15} in Fig.~\ref{koi12rvfig}. For simplicity we assumed a circular orbit for Kepler-448~b. \cite{Bourrier15} also performed a fit without a constraint on the eccentricity, and found a $3\sigma$ upper limit on the eccentricity of 0.72. For this fit they found a $3\sigma$ upper mass limit of $<10\,M_J$.

We simultaneously fit the Doppler tomographic, photometric, and radial velocity data using \texttt{emcee}. The MCMC used 16 parameters: $P$, $T_0$, $R_P/R_{\star}$, $a/R_{\star}$, $b$, $\lambda$, $v\sin i_{\star}$, the radial velocity semi-amplitude $K$ and velocity offset $\gamma$, the intrinsic (Gaussian) stellar line width, the central velocity and depth of the contaminating solar line, and two limb darkening parameters each for the Doppler tomographic and photometric datasets. We set Gaussian priors only upon the limb darkening parameters and set uniform priors on the other parameters (although with a cut-off of $K>0$, in addition to those described earlier). 

\begin{figure}
\epsscale{1.15}
\plotone{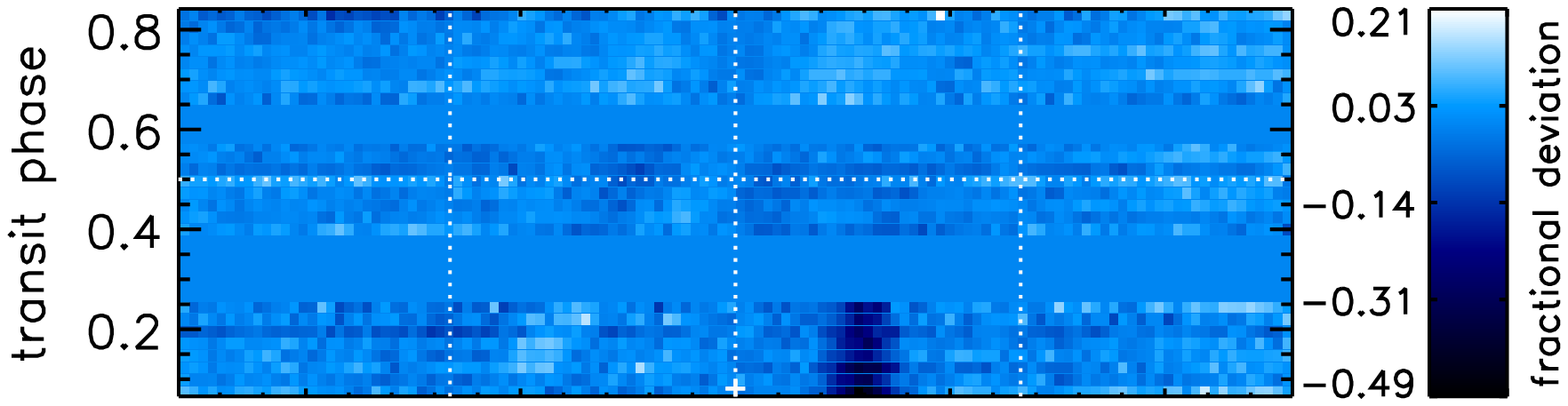}\\
\plotone{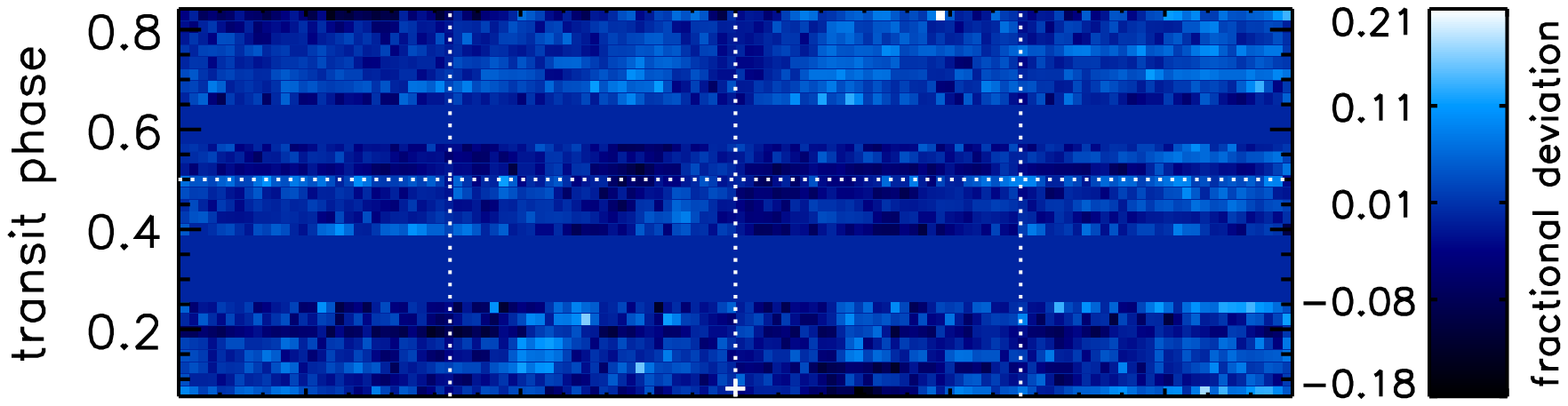}\\
\plotone{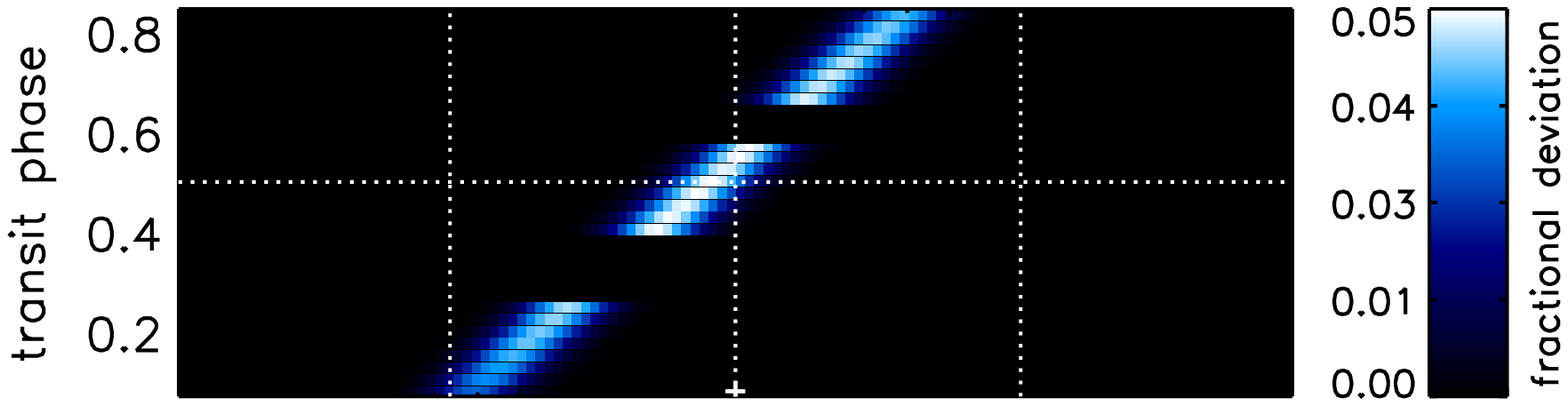}\\
\plotone{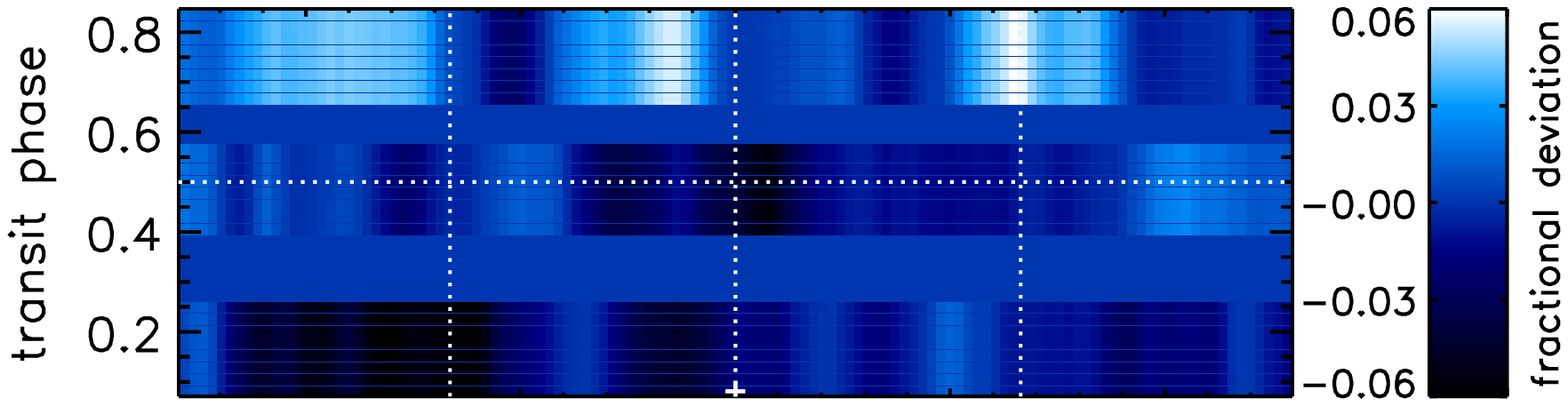}\\
\plotone{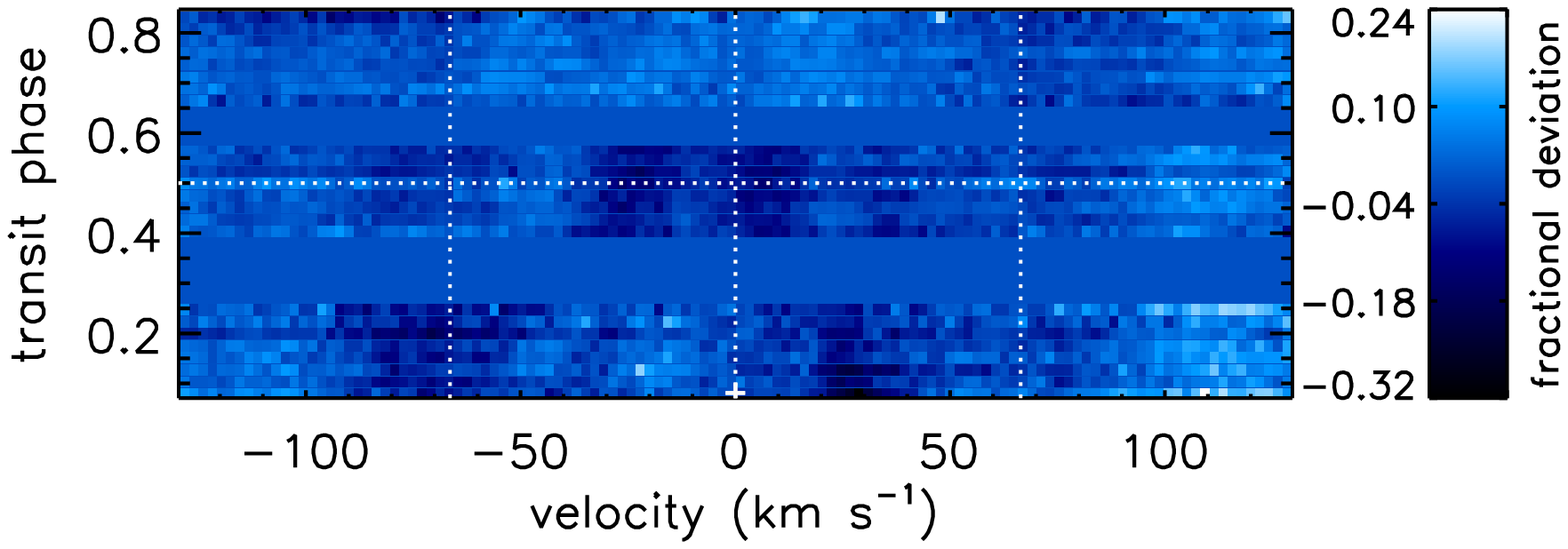}\\
\caption{Time series line profile residuals for Kepler-448~b. Top: the raw time series line profile residuals, showing the contaminating line due to moonlight. Top middle: with the best-fit model of the moonlight removed. The transit signature is the bright streak moving from lower left to upper right. The significant systematics are likely line profile perturbations caused by the same starspots responsible for the photometric variability of Kepler-448. Flat blue regions denote parts of the transit where we do not have observations. Middle: best-fitting transit model. Bottom middle: best-fit Gaussian process regression model. Bottom: data with the best-fit transit, moonlight, and Gaussian process regression models subtracted. Notation on this figure is the same as on Fig.~\ref{hatp41dt}. \label{koi12dt}} 
\end{figure}

We show the time series line profiles residuals for Kepler-448~b in Fig.~\ref{koi12dt}, and the best-fit models for the photometric and radial velocity data in Figs.~\ref{koi12trlc} and \ref{koi12rvfig}, respectively. Using the MCMC, we measured a spin-orbit misalignment of $\lambda = -7.1_{-2.8}^{+4.2\circ}$. At face value, this is $3.8\sigma$ discrepant from the value of $12.5_{-2.9}^{+3.0 \circ}$ found by \cite{Bourrier15}; however, a visual inspection of our best-fit model line profile perturbation and that of \cite{Bourrier15} shows that in both cases the line profile perturbation is located at negative velocities at mid-transit, whereas for $\lambda > 0$ the perturbation should be at positive velocities at this time. This suggests that there is actually no disagreement between the results, and that \cite{Bourrier15} used a different sign convention for $\lambda$. 
Nonetheless, given the long orbital period and consequently small tidal damping for Kepler-448~b, the small value of $\lambda$ suggests that the planet likely formed in and migrated through a well-aligned protoplanetary disk. 
As such, the presence of additional objects in the system is not required to have driven the migration of Kepler-448 b. There are no known stellar companions to Kepler-448 \citep{LilloBox14,Kraus16}, and no significant transit timing variations were seen in the {\it Kepler} data \citep{Mazeh13}, but long-term radial velocity monitoring to find additional planetary companions is unfeasible due to the rapid stellar rotation.

The system parameters that we measured are listed in the bottom section of Table~\ref{K448table}. These are generally compatible with those found by \cite{Bourrier15}, with the sole exception of $\lambda$ discussed above. We found a radial velocity semi-amplitude of $K=0.34^{+0.40}_{-0.29}$ km s$^{-1}$ \citep[corresponding to a planetary mass of $M_P=5.6_{-4.8}^{+6.6}\,M_J$, assuming the stellar mass from][]{Bourrier15}; however, given that these values differ from zero at a level of only $1.2\sigma$, we cannot claim a detection of the radial velocity reflex motion. We therefore instead quote $3\sigma$ upper limits of $K<1.55$ km s$^{-1}$ and $M_P<25.7\,M_J$.

\begin{deluxetable*}{lccc}
\tabletypesize{\footnotesize}
\tablecolumns{4}
\tablewidth{0pt}
\tablecaption{System Parameters of Kepler-448 \label{K448table}}
\tablehead{
 \colhead{Parameter} & \colhead{This Work} & \colhead{\cite{Bourrier15}} & \colhead{Prior} 
}

\startdata
Stellar Parameters & & \\
$T_{\mathrm{eff}}$ (K) & \ldots & $6820 \pm 120$ & \ldots \\
$M_{\star}\,(M_{\odot})$ & \ldots & $1.452 \pm 0.093$ & \ldots \\
$R_{\star}\,(R_{\odot})$ & \ldots & $1.63 \pm 0.15$ & \ldots \\
$P_{\mathrm{rot}}$ (days) & $1.27 \pm 0.11$ & \ldots & \ldots \\
\hline
Measured Parameters & &  \\
$P$ (days) & $17.85523216^{+0.00000057}_{-0.00000051}$ & $17.8552332 \pm 0.0000010$ & $\mathcal{U}(0,\infty)$ \\
$T_0$ (BJD) & $2454979.596045 \pm 0.000024$ & $2454979.59601 \pm 0.00005$ & $\mathcal{U}(-\infty,\infty)$ \\
$R_P/R_{\star}$ & $0.090537^{+0.000037}_{-0.000028}$ & $0.09049 \pm 0.00008$ & $\mathcal{U}(0,1)$ \\
$a/R_{\star}$ & $18.82^{+0.021}_{-0.022}$ & $18.84 \pm 0.04$ &  $\mathcal{U}(0,\infty)$ \\
$b$ & $0.3661^{+0.0031}_{-0.0027}$ & $0.362 \pm 0.007$ & $\mathcal{U}(-(1+R_P/R_{\star}),1+R_P/R_{\star})$ \\
$\lambda\,(^{\circ})$ & $-7.1^{+4.2}_{-2.8}$ & $12.5_{-2.9}^{+3.0}$ & $\mathcal{U}(-180,180)$ \\
$v\sin i_{\star}$ (km s$^{-1}$) & $66.43_{-0.95}^{+1.00}$ & $60.0_{-0.8}^{+0.9}$  &  $\mathcal{U}(0,\infty)$\\
$K$ (km s$^{-1}$) & $<1.55$ ($3\sigma$) & $<0.51$ ($3\sigma$) & $\mathcal{U}(0,\infty)$ \\
intrinsic line width (km s$^{-1}$) & $7.8^{+2.1}_{-1.6}$ & \ldots & $\mathcal{U}(0,\infty)$ \\
\hline
Derived Parameters & &  \\
$M_P\,(M_J)$\tablenotemark{a} & $<25.7$ ($3\sigma$) & $<8.7$ ($3\sigma$) & \ldots \\
$R_P\,(R_J)$\tablenotemark{a} & $1.44 \pm 0.13$ & $1.44 \pm 0.13$  & \ldots \\
\enddata

\tablecomments{Measured parameters are those that we measured directly with the MCMC, while Derived Parameters are calculated analytically from the Measured Parameters. The Prior column lists the type of prior used for each parameter in the MCMC. Notation: $\mathcal{U}(x,y)$: uniform prior between $x$ and $y$. Parameters with no prior type listed were not used in our MCMC and are quoted here for informational purposes only.}
\tablenotetext{a}{To calculate the planetary mass and radius we assumed the stellar parameters found by \cite{Bourrier15}, as we did not calculate our own values for these parameters.}

\end{deluxetable*}

\subsection{The Rotation Period of Kepler-448}\label{rotationsection}

\begin{figure}
\epsscale{1.2}
\plotone{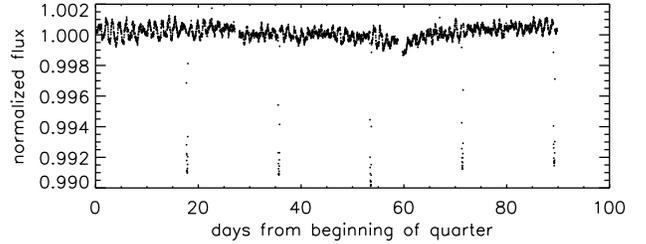}
\caption{A representative \emph{Kepler} light curve of Kepler-448, showing the rotational variability and the planetary transits. These data are from Quarter 6 of the \emph{Kepler} mission. The data are short cadence, which are broken into three shorter intervals for each quarter, hence the discontinuities at $\sim28$ and $\sim60$ days after the start of the quarter. \label{koi12lcfig}} 
\end{figure}

The most obvious features in the \emph{Kepler} light curve of Kepler-448 (Fig.~\ref{koi12lcfig}) are the planetary transits (period 17.9 days) and a quasi-sinusoidal modulation of maximum amplitude $\sim 0.2$\%, which changes in amplitude and phase 
on timescales of tens of days. We attribute this signal to rotational modulation, as spots on the stellar surface move in and out of view; the variability is too irregular to be due to stellar pulsations.

In order to investigate this signal further, we followed the methodology of \cite{McQuillan13}, who developed tools to investigate rotational modulation in \emph{Kepler} data. For each \emph{Kepler} quarter, we calculated the autocorrelation function (ACF) of the long cadence light curve. Each ACF shows a peak at a period of $\sim1.3$ days, plus a series of peaks at longer periods spaced at intervals of approximately 1.3 days. In principle, any peak may be the rotation period, with peaks at shorter periods due to spots on opposite sides of the star resulting in periodicity on half the true rotation period. Nonetheless, we identify the 1.3 day peak as the rotation period for the following reason. The rotation period, equatorial velocity, and stellar radius are related by $P_{\mathrm{rot}}=2\pi R_{\star}/v_{\mathrm{eq}}$, where the equatorial velocity is $v_{\mathrm{eq}}=(v\sin i_{\star})/\sin i_{\star}$. Given measured values of $P_{\mathrm{rot}}$ and $v\sin i_{\star}$, we therefore have $R_{\star}\sin i_{\star}=P_{\mathrm{rot}}v\sin i_{\star}/2\pi$. The minimum possible value for $R_{\star}$ will occur for $i_{\star}=90^{\circ}$, with smaller values of $i_{\star}$ requiring larger values of $R_{\star}$. A rotation period of 1.3 days and a $v\sin i_{\star}$ of 66 km s$^{-1}$ would thus give a minimum stellar radius of 1.7 $R_{\odot}$, broadly consistent with the stellar parameters derived from spectroscopy \citep{Bourrier15}. If the second peak, at $\sim2.6$ days, instead corresponded to the rotation period, this would require a minimum stellar radius of 3.5 $R_{\odot}$, inconsistent with the known stellar parameters. This also implies that $\sin i_{\star}\sim1$ (i.e., $i_{\star}\sim90^{\circ}$), and, together with the well-aligned orbit as projected onto the sky ($\lambda\sim0^{\circ}$), indicates that the full three-dimensional spin-orbit misalignment is small, $\psi\sim0^{\circ}$.

For each of the eight quarters with a well-behaved ACF (i.e., with distinct, approximately equally spaced peaks in the 1-5 day range and without an excess of power at short periods), we measured the rotation period from the location of the peak near 1.3 days, and the uncertainty in the period from the half width half maximum of the peak, following \cite{McQuillan13}. Our reported rotation period is the mean of the periods measured for each of these quarters. This resulted in a period of $P_{\mathrm{rot}}=1.27 \pm 0.11$ days.

\cite{McQuillan13}, however, noted that the ACF method is not necessarily reliable for rotation periods below 7 days. In order to double-check our results, we also calculated the Lomb-Scargle periodogram \citep{Lomb76,Scargle82,ZechmeisterKurster09} of the \emph{Kepler} light curve for each quarter. For the same quarters that we used to measure the period from the ACF, the mean period measured from the Lomb-Scargle periodogram was 1.26 days, in good agreement with the value found by the ACF method. Additionally, this rotation period is broadly consistent with the values of 1.245 days found by \cite{McQuillan13b} and 1.23 days found by \cite{Mazeh15}. We thus adopt $P_{\mathrm{rot}}=1.27 \pm 0.11$ days as the rotation period of Kepler-448.

\section{Conclusions}

We have presented Doppler tomographic observations of three giant planets--the hot Jupiters HAT-P-41~b and WASP-79~b, and the warm Jupiter Kepler-448~b--and used these to measure the sky-projected angle between the stellar spin and planetary orbital angular momentum vectors, $\lambda$. For HAT-P-41 b we measured $\lambda = -22.1_{-6.0}^{+0.8 \circ}$, suggesting a somewhat misaligned but prograde orbit, as is typical for hot Jupiters around stars above the Kraft break. 

We also reanalyzed the data that \cite{Addison13} obtained on WASP-79~b using Doppler tomographic methodology. We obtained a measurement of $\lambda=-99.1_{-3.9}^{+4.1 \circ}$, which is consistent with, but significantly more precise than, the value of $\lambda=-106^{+19 \circ}_{-13}$ found by \cite{Addison13}. This demonstrates the power of Doppler tomography, especially for mildly rapidly rotating stars like WASP-79 where the achievable radial velocity precision begins to be degraded by significant rotational broadening. Similar results have recently been found by \cite{Brown16}, who analyzed several datasets with both radial velocity Rossiter-McLaughlin and Doppler tomographic methodology, and in all cases Doppler tomography returned more precise measurements of $\lambda$. In addition, Doppler tomography is not as susceptible to effects like convective blueshift and variations in convection across the stellar disk \citep[e.g.,][]{Cegla16}, obscuration of stellar active regions \citep{Oshagh16}, and night-to-night velocity offsets that can introduce systematic uncertainties into radial velocity Rossiter-McLaughlin analysis. For WASP-79 we also used our results to argue that the main sequence solution presented by \cite{Smalley12} is the correct solution for the system parameters.

Finally, we performed a full analysis of the parameters of the warm Jupiter Kepler-448~b. We found a nearly well-aligned orbit ($\lambda=-7.1^{+4.2 \circ}_{-2.8}$), a value that is likely consistent with that previously found by \cite{Bourrier15}; our other measured system parameters are in agreement with those found by \cite{Bourrier15}. We also used the rotational modulation in the {\it Kepler} light curve to measure a stellar rotation period of $P_{\mathrm{rot}}=1.27 \pm 0.11$ days. Given the stellar parameters from \cite{Bourrier15} and the measured value of $v\sin i_{\star}$, this implies that $\sin i_{\star}\sim1$ and that the orbit is well-aligned in three-dimensional space, $\psi\sim0^{\circ}$.

All three stars observed in this work are above the Kraft break, and none of them have additional confirmed objects in their systems that could have driven the migration of the planet. WASP-79~b, with its nearly-polar orbit, and HAT-P-41~b, with its somewhat misaligned orbit, clearly follow the trend that hot Jupiters around hot stars have a wide range of spin-orbit misalignments \citep{Winn10}.

The picture is also less clear for Kepler-448 b; since the number of warm Jupiters with measured spin-orbit misalignments is small, no pattern has yet emerged. Indeed, Kepler-448 b is the \emph{only} warm Jupiter with a host star above the Kraft break and a measured spin-orbit misalignment. 
That it is aligned \citep[or has only a small misalignment;][]{Bourrier15} could provide a tantalizing hint that the spin-orbit misalignment distribution of warm Jupiters is different than that of hot Jupiters; the $\lambda$ distribution of hot Jupiters around hot stars is consistent with isotropic \citep{Albrecht13}, and the odds that a single planet drawn from this distribution would be aligned are small. If true, this would disfavor misalignment generation models like that of \cite{Rogers12} and \cite{Batygin12}, which predict similar spin-orbit misalignment distributions for hot and warm Jupiters around hot stars. On the other hand, \cite{Crouzet16} suggested that hot Jupiters around the hottest stars ($T_{\mathrm{eff}}>6700$ K, like Kepler-448) may on average have smaller misalignments than those around stars with $6250$ K$<T_{\mathrm{eff}}<6700$ K. Obviously, more spin-orbit misalignment measurements for warm Jupiters around hot stars are necessary to draw any solid conclusions on this point.

\vspace{12pt}

Thanks to Erik Brugamyer, Andrew Collier Cameron, Michael Endl, and Alexandre Santerne for useful discussions. We thank the HET Resident Astronomers and Telescope Operators who obtained our data on HAT-P-41~b and Kepler-448~b, and the anonymous referee for helpful comments that increased the quality of the paper. 

M.C.J. was supported in part by a NASA Earth and Space Science Fellowship under grant NNX12AL59H, and by the NASA K2 Guest Observer Program under grants NNX15AV58G (Cycle 1) and NNX16AE58G (Cycle 2) to the University of Texas at Austin. 

This paper includes data taken at The McDonald Observatory of The University of Texas at Austin. The Hobby-Eberly Telescope (HET) is a joint project of the University of Texas at Austin, the Pennsylvania State University, Stanford University, Ludwig-Maximilians-Universit\"at M\"unchen, and Georg-August-Universit\"at G\"ottingen. The HET is named in honor of its principal benefactors, William P. Hobby and Robert E. Eberly. This work has made use of the VALD database, operated at Uppsala University, the Institute of Astronomy RAS in Moscow, and the University of Vienna.

\vspace{5mm}
\facilities{HET(HRS), AAT(UCLES)}
\software{IDL, Python, IRAF, MATLAB, emcee \citep{ForemanMackey13}, george \citep{Ambikasaran14}, Batman \citep{Kreidberg15}, JKTLD \citep{Southworth15}, PyKE \citep{StillBarclay12}}

\end{document}